\documentclass[aps,prx,twocolumn,superscriptaddress]{revtex4-1}%
\usepackage{graphics}
\usepackage{amsmath}
\usepackage{graphicx}
\usepackage{amsfonts}
\usepackage{amssymb}
\usepackage{float}
\usepackage{longtable}
\usepackage{epsfig}
\usepackage{latexsym}
\usepackage{theorem}
\usepackage{bbm}
\usepackage{bm}
\usepackage{verbatim}
\usepackage{psfrag}
\usepackage{xcolor}

\newtheorem{theorem}{Theorem}
\newtheorem{lemma}[theorem]{Lemma}

\begin{document}
\title{Fundamental limits to quantum channel discrimination}
\author{Stefano Pirandola}
\email{pirs@mit.edu, stefano.pirandola@york.ac.uk}
\affiliation{Research Laboratory of Electronics, Massachusetts
Institute of Technology, Cambridge, Massachusetts 02139, USA}
\affiliation{Computer Science and York Centre for Quantum
Technologies, University of York, York YO10 5GH, UK}
\author{Riccardo Laurenza}
\affiliation{QSTAR, INO-CNR and LENS, Largo Enrico Fermi 2, 50125 Firenze, Italy}
\author{Cosmo Lupo}
\affiliation{Department of Physics and Astronomy, University of Sheffield, Sheffield S3
7RH, UK}
\author{Jason L. Pereira}
\affiliation{Computer Science and York Centre for Quantum Technologies, University of York,
York YO10 5GH, UK}

\begin{abstract}
What is the ultimate performance for discriminating two arbitrary
quantum channels acting on a finite-dimensional Hilbert space?
Here we address this basic question by deriving a general and
fundamental lower bound. More precisely, we investigate the
symmetric discrimination of two arbitrary qudit channels by means
of the most general protocols based on adaptive
(feedback-assisted) quantum operations. In this general scenario,
we first show how port-based teleportation can be used to simplify
these adaptive protocols into a much simpler non-adaptive form,
designing a new type of teleportation stretching. Then, we prove
that the minimum error probability affecting the channel
discrimination cannot beat a bound determined by the Choi matrices
of the channels, establishing a general, yet computable formula
for quantum hypothesis testing. As a consequence of this bound, we
derive ultimate limits and no-go theorems for adaptive quantum
illumination and single-photon quantum optical resolution.
Finally, we show how the methodology can also be applied to other
tasks, such as quantum metrology, quantum communication and secret
key generation.

\end{abstract}
\maketitle

\section{Introduction}
Quantum hypothesis testing~\cite{QHT1} is a central area in quantum
information theory~\cite{Watrous,HolevoBOOK}, with many studies for both
discrete variable (DV)~\cite{NiCh} and continuous variable (CV)
systems~\cite{RMP}. A number of tools~\cite{QCB1,QCB2,QCB3,QHB1,QHB2} have
been developed for its basic formulation, known as quantum state
discrimination. In particular, since the seminal work of Helstrom in the
70s~\cite{QHT1}, we know how to bound the error probability affecting the
symmetric discrimination of two arbitrary quantum states. Remarkably, after
about 40 years, a similar bound is still missing for the discrimination of two
arbitrary quantum channels. There is a precise motivation for that: The main
problem in quantum channel discrimination
(QCD)~\cite{QCD1,QCD2,QCD3,QCD4,QCD6} is that the strategies involve an
optimization over the input states and the output measurements, and this
process may be adaptive in the most general case, so that feedback from the
output can be used to update the input.

Not only the ultimate performance of adaptive QCD\ is still
unknown due to the difficulty of handling feedback-assistance, but
it is also known that adaptiveness needs to be considered in QCD.
In fact, apart from the cases where two channels are
classical~\cite{HayaCLASS}, jointly programmable or
teleportation-covariant~\cite{PirCo,ReviewMETRO}, feedback may
greatly improve the discrimination. For instance, Ref.~\cite{Harrow}%
\ presented two channels which can be perfectly distinguished by
using feedback in just two adaptive uses, while they cannot be
perfectly discriminated by any number of uses of a block
(non-adaptive) protocol, where the channels are probed in an
identical and independent fashion. This suggests that the best
discrimination performance is not directly related to the diamond
distance~\cite{Diamond}, when computed over multiple copies of the
quantum channels.

In this work we finally fill this fundamental gap by deriving a
universal computable lower bound for the error probability
affecting the discrimination of two \textit{arbitrary} quantum
channels. To derive this bound we adopt a technique which reduces
an adaptive protocol over an arbitrary finite-dimensional quantum
channel into a block protocol over multiple copies of the
channel's Choi matrix. This is obtained by using port-based
teleportation (PBT)~\cite{PBT,PBT1,PBT2,Brau} for channel
simulation and suitably generalizing the technique of
teleportation stretching~\cite{PLOB,TQC,Uniform}. This\ reduction
is shown for adaptive protocols with any task (not just QCD). When
applied to QCD, it allows us to bound the ultimate error
probability by using the Choi matrices of the channels.

As a direct application, we bound the ultimate adaptive
performance of quantum
illumination~\cite{Qill0,Qill1,Qill2,Qill3,Qill4,Qill6,Qill7,Qill8}
and the ultimate adaptive resolution of any single-photon
diffraction-limited optical system, setting corresponding no-go
theorems for these applications. We then apply our result to
adaptive quantum metrology showing an ultimate bound which has an
asymptotic Heisenberg scaling. As an example, we also study the
adaptive discrimination of amplitude damping channels, which are
the most difficult channels to be simulated. Finally, other
implications are for the two-way assisted capacities of quantum
and private communications.

\section{Results}

\subsection{Adaptive protocols}

\begin{figure}[ptbh]
\begin{center}
\vspace{-2cm} \includegraphics[width=0.49\textwidth]{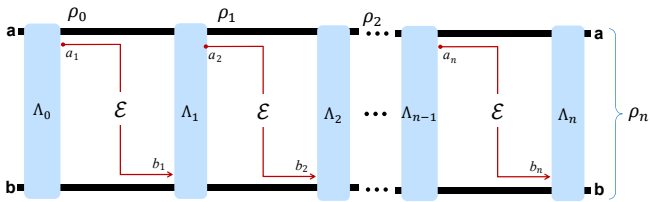} \vspace{-2.4cm}
\end{center}
\caption{General structure of an adaptive quantum protocol, where
channel uses $\mathcal{E}$ are interleaved by QOs $\Lambda$'s. See
text for more details.}
\label{figada}%
\end{figure}

Let us formulate the most general adaptive protocol over an arbitrary quantum
channel $\mathcal{E}$ defined between Hilbert spaces of dimension $d$ (more
generally, this can be taken as the dimension of the input space). We first
provide a general description and then we specify the protocol to the task of
QCD. A general adaptive protocol involves an unconstrained number of quantum
systems which may be subject to completely arbitrary quantum operations (QOs).
More precisely, we may organize the quantum systems into an input register
$\mathbf{a}$ and an output register $\mathbf{b}$, which are prepared in an
initial state $\rho_{0}$ by applying a QO $\Lambda_{0}$ to some fundamental
state of $\mathbf{a}$ and $\mathbf{b}$. Then, a system $a_{1}$ is picked from
the register $\mathbf{a}$ and sent through the channel $\mathcal{E}$. The
corresponding output $b_{1}$ is merged with the output register $b_{1}%
\mathbf{b\rightarrow b}$. This is followed by another QO $\Lambda_{1}$ applied
to $\mathbf{a}$ and $\mathbf{b}$. Then, we send a second system $a_{2}%
\in\mathbf{a}$ through $\mathcal{E}$ with the output $b_{2}$ being
merged again $b_{2}\mathbf{b\rightarrow b}$ and so on. After $n$
uses, the registers will be in a state $\rho_{n}$ which depends on
$\mathcal{E}$ and the sequence of QOs
$\{\Lambda_{0},\Lambda_{1},\ldots,\Lambda_{n}\}$ defining the
adaptive protocol $\mathcal{P}_{n}$ with output state $\rho_{n}$
(see Fig.~\ref{figada}).

In a protocol of quantum communication, the registers belong to remote users
and, in absence of entanglement-assistance, the QOs are local operations (LOs)
assisted by two-way classical communication (CC), also known as adaptive
LOCCs. The output is generated in such a way to approximate some target
state~\cite{PLOB}. In a protocol of quantum channel estimation, the channel is
labelled by a continuous parameter $\mathcal{E}=\mathcal{E}_{\theta}$ and the
QOs include the use of entanglement across the registers. The output state
will encode the unknown parameter $\rho_{n}=\rho_{n}(\theta)$, which is
detected and the outcome processed into an optimal estimator~\cite{PirCo}.
Here, in a protocol of binary and symmetric QCD, the channel is labelled by a
binary digit, i.e., $\mathcal{E}=\mathcal{E}_{u}$ where $u\in\{0,1\}$ has
equal priors. The QOs are generally entangled and they generate an output
state encoding the information bit, i.e., $\rho_{n}=\rho_{n}(u)$.

The output state $\rho_{n}(u)$ of an adaptive discrimination protocol
$\mathcal{P}_{n}$\ is finally detected by an optimal positive-operator valued
measure (POVM). For binary discrimination, this is the Helstrom POVM, which
leads to the conditional error probability
\begin{equation}
p(\mathcal{E}_{0}\neq\mathcal{E}_{1}|\mathcal{P}_{n})=\frac{1-D\left[
\rho_{n}(0),\rho_{n}(1)\right]  }{2}, \label{protPROB}%
\end{equation}
where $D(\rho,\sigma):=||\rho-\sigma||/2$ is the trace distance~\cite{NiCh}.
The optimization over all discrimination protocols $\mathcal{P}_{n}$ defines
the minimum error probability affecting the $n$-use adaptive discrimination of
$\mathcal{E}_{0}$ and $\mathcal{E}_{1}$, i.e., we may write%
\begin{equation}
p_{n}(\mathcal{E}_{0}\neq\mathcal{E}_{1}):=\inf_{\mathcal{P}_{n}%
}~p(\mathcal{E}_{0}\neq\mathcal{E}_{1}|\mathcal{P}_{n}). \label{protINF}%
\end{equation}

This is generally less than the $n$-copy diamond distance between the two
channels $\mathcal{E}_{0}^{\otimes n}$ and $\mathcal{E}_{1}^{\otimes n}$%
\begin{equation}
p_{n}(\mathcal{E}_{0}\neq\mathcal{E}_{1})\leq\frac{1-\frac{1}{2}%
||\mathcal{E}_{0}^{\otimes n}-\mathcal{E}_{1}^{\otimes n}||_{\diamond}}{2},
\label{eqDD}%
\end{equation}
where~\cite{Watrous}
\begin{equation}
||\mathcal{E}_{0}^{\otimes n}-\mathcal{E}_{1}^{\otimes n}||_{\diamond}%
:=\sup_{\rho_{ar}}||\mathcal{E}_{0}^{\otimes n}\otimes\mathcal{I}(\rho
_{ar})-\mathcal{E}_{1}^{\otimes n}\otimes\mathcal{I}(\rho_{ar})||,
\end{equation}
with $\mathcal{I}$ being an identity map acting on a reference
system $r$. The upper bound in Eq.~(\ref{eqDD}) is achieved by a
non-adaptive protocol, where an (optimal) input state $\rho_{ar}$
is prepared and its $a$-parts transmitted through
$\mathcal{E}_{u}^{\otimes n}$. Note that Eq.~(\ref{eqDD}) is very
difficult to compute, which is why we usually compute larger but
simpler single-letter upper bounds such as
\begin{equation}
p_{n}(\mathcal{E}_{0}\neq\mathcal{E}_{1})\leq\frac{F(\rho_{\mathcal{E}_{0}%
},\rho_{\mathcal{E}_{1}})^{n}}{2}, \label{fidSIM}%
\end{equation}
where $F$ is the fidelity between the Choi matrices,
$\rho_{\mathcal{E}_{0}}$ and $\rho_{\mathcal{E}_{1}}$, of the two
channels.

Our question is: Can we complete Eq.~(\ref{eqDD}) with a
corresponding lower bound? Up to today this has been only proven
for jointly-programmable channels, i.e., channels
$\mathcal{E}_{0}$ and
$\mathcal{E}_{1}$ admitting a simulation $\mathcal{E}_{u}(\rho)=\mathcal{S}%
(\rho\otimes\pi_{u})$ with a trace-preserving QO $\mathcal{S}$ and different
program states $\pi_{0}$ and $\pi_{1}$. In this case, we have $p_{n}%
\geq\lbrack1-D(\pi_{0}^{\otimes n},\pi_{1}^{\otimes n})]/2$~\cite{PirCo}. In
particular, this is true if the channels are jointly teleportation-covariant,
so that $\mathcal{S}$ becomes teleportation and the program state is a Choi
matrix $\rho_{\mathcal{E}_{u}}$. For these channels, Ref.~\cite{PirCo} found
that Eq.~(\ref{eqDD}) holds with an equality and we may write $||\mathcal{E}%
_{0}^{\otimes n}-\mathcal{E}_{1}^{\otimes n}||_{\diamond}=||\rho
_{\mathcal{E}_{0}}^{\otimes n}-\rho_{\mathcal{E}_{1}}^{\otimes n}||$. More
precisely, the question to ask is therefore the following: Can we establish a
\textit{universal} lower bound for $p_{n}(\mathcal{E}_{0}\neq\mathcal{E}_{1})$
which is valid for\textit{ arbitrary} channels? As we show here, this is
possible by resorting to a more general (multi-program) simulation of the
channels, i.e., of the type $\mathcal{S}(\rho\otimes\pi_{u}^{\otimes M})$.

\begin{figure*}[ptbh]
\begin{center}
\vspace{-4.6cm} \includegraphics[width=1.05\textwidth]{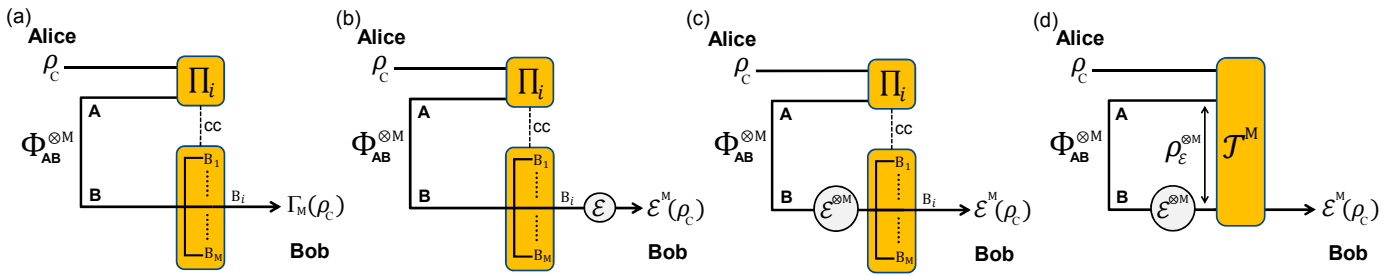} \vspace{-0.4cm}
\vspace{-5.2cm}
\end{center}
\caption{From port-based teleportation (PBT) to Choi-simulation of a quantum
channel (see also Ref.~\cite{PBT}). \textbf{(a)} Schematic representation of
the PBT protocol. Alice and Bob share an $M\times M$ qudit state which is
given by $M$ maximally-entangled states $\Phi_{\mathbf{AB}}^{\otimes M}$. To
teleport an input qubit state $\rho_{C}$, Alice applies a suitable POVM
$\{\Pi_{i}\}$ to the input qubit $C$ and her $\mathbf{A}$ qubits. The outcome
$i$ is communicated to Bob, who selects the $i$-th among his $\mathbf{B}$
qubits (tracing all the others). The performance does not depend on the
specific \textquotedblleft port\textquotedblright\ $i$ selected and the
average output state is given by $\Gamma_{M}(\rho_{C})$ where $\Gamma_{M}$ is
the PBT channel. The latter reduces to the identity channel in the limit of
many ports $M\rightarrow\infty$. \textbf{(b)} Suppose that Bob applies a
quantum channel $\mathcal{E}$ on his teleported output. This produces the
output state $\mathcal{E}^{M}(\rho_{C})$ of Eq.~(\ref{effective}). For large
$M$, one has $\mathcal{E}^{M}\rightarrow\mathcal{E}$ in diamond norm.
\textbf{(c)} Equivalently, Bob can apply $\mathcal{E}^{\otimes M}$ to all his
qubits $\mathbf{B}$ in advance to the CC from Alice. After selection of the
port, this will result in the same output as before. \textbf{(d)} Now note
that Alice's LO and Bob's port selection form a global LOCC $\mathcal{T}^{M}$
(trace-preserving by averaging over the outcomes). This is applied to a
tensor-product state $\rho_{\mathcal{E}}^{\otimes M}$ where $\rho
_{\mathcal{E}}$ is the Choi matrix of the original channel $\mathcal{E}$. Thus
the approximate channel $\mathcal{E}^{M}$ is simulated by applying
$\mathcal{T}^{M}$ to $\rho_{C}\otimes\rho_{\mathcal{E}}^{\otimes M}$ as in
Eq.~(\ref{LOCCsim}).}%
\label{fig1}%
\end{figure*}

\subsection{PBT and simulation of the identity}

Let us describe the protocol of PBT with qudits of arbitrary dimension
$d\geq2$. More technical details can be found in the original
proposals~\cite{PBT1,PBT2}. The parties exploit two ensembles of $M\geq2$
qudits, i.e., Alice has $\mathbf{A}:=\{A_{1},\ldots,A_{M}\}$ and Bob has
$\mathbf{B}:=\{B_{1},\ldots,B_{M}\}$ representing the output \textquotedblleft
ports\textquotedblright. The generic $i$th pair $(A_{i},B_{i})$ is prepared in
a maximally-entangled state, so that we have the global state%
\begin{equation}
\Phi_{\mathbf{AB}}^{\otimes M}=\bigotimes_{i=1}^{M}|\Phi\rangle_{i}\langle
\Phi|,~~|\Phi\rangle_{i}:=d^{-1/2}\sum_{k}\left\vert k\right\rangle _{A_{i}%
}\otimes\left\vert k\right\rangle _{B_{i}}. \label{nonOPT}%
\end{equation}
To teleport the state of a qudit $C$, Alice performs a joint measurement on
$C$ and her ensemble $\mathbf{A}$. This is a POVM $\{\Pi_{C\mathbf{A}}%
^{i}\}_{i=1}^{M}$ with $M$ possible outcomes (see
Refs.~\cite{PBT1,PBT2} for the details). In the standard protocol
considered here, this POVM is a square root measurement (known to
be optimal in the qubit case). Once Alice communicates the outcome
$i$ to Bob, he discards all the ports but the $i$th one, which
contains the teleported state (see Fig.~2a).

The measurement outcomes are equiprobable and independent of the input, and
the output state is invariant under permutation of the ports (this can be
understood by the fact that the scheme is invariant under permutation of the
Bell states and, therefore, of the ports). Averaging over the outcomes, we
define the teleported state $\rho_{B}^{M}=\Gamma_{M}(\rho_{C})$, where
$\Gamma_{M}$ is the corresponding PBT\ channel. Explicitly, this channel takes
the form
\begin{equation}
\Gamma_{M}(\rho_{C})=\sum_{i=1}^{M}\mathrm{Tr}_{\mathbf{A}\bar{B_{i}}C}%
[\Pi_{C\mathbf{A}}^{i}\left(  \rho_{C}\otimes\Phi_{\mathbf{AB}}^{\otimes
M}\right)  ],
\end{equation}
where $\text{Tr}_{\bar{B_{i}}}$ denotes the trace over all ports $\mathbf{B}$
but $B_{i}$.

As shown in Ref.~\cite{PBT1}, the standard protocol gives a
depolarizing channel~\cite{NiCh} whose probability $\xi_{M}$
decreases to zero for increasing number of ports $M$. Therefore,
in the limit of many ports $M\gg1$, the $M$-port PBT channel
$\Gamma_{M}$ tends to an identity channel $\mathcal{I}$, so that
Bob's output becomes a perfect replica of Alice's input. Here we
prove a stronger result in terms of channel uniform
convergence~\cite{TQC,Uniform}. In fact, for any $M$, we show that
the simulation error, expressed in terms of the diamond distance
between $\Gamma_{M}$ and $\mathcal{I}$, is one-to-one with the
entanglement fidelity of the PBT channel $\Gamma_{M}$. In turn,
this result allows us to write a simple upper bound for this
error. Moreover, we can fully characterize the simulation error
with an exact analytical expression for qubits (see Methods for
the proof, with further details being given in Supplementary
Section~I).

\begin{lemma}
\label{lemmaPBT}In arbitrary (finite) dimension $d$, the diamond distance
between the $M$-port PBT channel $\Gamma_{M}$ and the identity channel
$\mathcal{I}$ satisfies%
\begin{equation}
\delta_{M}:=||\mathcal{I}-\Gamma_{M}||_{\diamond}=2[1-f_{e}(\Gamma_{M})],
\label{firstEQQ}%
\end{equation}
where $f_{e}(\Gamma_{M}):=\langle\Phi|[\mathcal{I}\otimes\Gamma_{M}%
(|\Phi\rangle\langle\Phi|)]|\Phi\rangle$ is the entanglement fidelity of
$\Gamma_{M}$. This gives the upper bound%
\begin{equation}
\delta_{M}\leq2d(d-1)M^{-1}~. \label{kbound}%
\end{equation}
More precisely, we can write the exact result
\begin{equation}
\delta_{M}=\frac{2\left(  d^{2}-1\right)  }{d^{2}}\xi_{M}, \label{exactBB}%
\end{equation}
where $\xi_{M}$ is the depolarizing probability of the PBT\
channel $\Gamma_{M}$. For qubits ($d=2$), the \textquotedblleft
PBT\ number\textquotedblright\ $\xi_{M}$ has the closed analytical
expression
\begin{align}
\xi_{M}  &  =\frac{1}{3}\frac{M+2}{2^{M-1}}+\frac{1}{3}\sum_{s=s_{min}%
}^{(M-1)/2}\frac{s(s+1)}{2^{M-4}}\binom{M}{\frac{M-1}{2}-s}\times\nonumber\\
&  \frac{\left(  M+2\right)  -\sqrt{\left(  M+2\right)
^{2}-\left( 2s+1\right)  ^{2}}}{\left(  M+2\right)  ^{2}-\left(
2s+1\right)  ^{2}}, \label{PBTnumbers}
\end{align}
where $s_{min}=1/2$ for even $M$ and $0$ for odd $M$.
\end{lemma}

\subsection{General channel simulation via PBT}

Let us discuss how PBT can be used for channel simulation. This was first
shown in Ref.~\cite{PBT}\ where PBT was introduced as a possible design for a
programmable quantum gate array~\cite{Array}. As depicted in Fig.~\ref{fig1}b,
suppose that Bob applies an arbitrary channel $\mathcal{E}$ to the teleported
output, so that Alice's input $\rho_{C}$ is subject to the approximate
channel
\begin{equation}
\mathcal{E}^{M}(\rho_{C}):=\mathcal{E}\circ\Gamma_{M}(\rho_{C}).
\label{effective}%
\end{equation}
Note that the port selection commutes with $\mathcal{E}$, because the POVM
acts on a different Hilbert space~\cite{PBT}. Therefore, Bob can equivalently
apply $\mathcal{E}$ to each port before Alice's CC, i.e., apply $\mathcal{E}%
^{\otimes M}$ to his $\mathbf{B}$ qudits before selecting the output port, as
shown in Fig.~\ref{fig1}c. This leads to the following simulation for the
approximate channel%
\begin{equation}
\mathcal{E}^{M}(\rho_{C})=\mathcal{T}^{M}(\rho_{C}\otimes\rho_{\mathcal{E}%
}^{\otimes M})~, \label{LOCCsim}%
\end{equation}
where $\mathcal{T}^{M}$ is a trace-preserving LOCC and $\rho_{\mathcal{E}}$ is
the channel's Choi matrix (see Fig.~\ref{fig1}d). By construction, the
simulation LOCC $\mathcal{T}^{M}$ is universal, i.e., it does not depend on
the channel $\mathcal{E}$. This means that, at fixed $M$, the channel
$\mathcal{E}^{M}$ is fully determined by the program state $\rho_{\mathcal{E}%
}$. One can bound the accuracy of the simulation. From Eq.~(\ref{effective})
and the monotonicity of the diamond norm, we get
\begin{equation}
||\mathcal{E}-\mathcal{E}^{M}||_{\diamond}\leq\delta_{M},
\label{simulationERROR}%
\end{equation}
where $\delta_{M}$ is the simulation error in Eq.~(\ref{kbound}), with the
dimension $d$ being the one of the input Hilbert space. It is worth to remark
that, while the simulation in Eq.~(\ref{LOCCsim}) relies on a number of copies
of the channel's Choi matrix, it can be applied to an arbitrary quantum
channel $\mathcal{E}$ without the condition of teleportation
covariance~\cite{PLOB}.


\begin{figure*}[ptb]
\begin{center}
\vspace{-4.6cm} \includegraphics[width=0.99\textwidth]{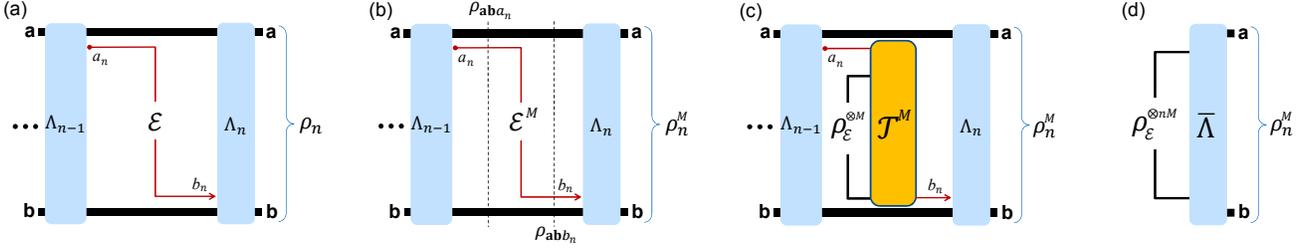} \vspace{-5cm}
\end{center}
\caption{Port-based teleportation stretching of a generic adaptive protocol
over a quantum channel $\mathcal{E}$. This channel is fixed in quantum/private
communication, while it is unknown and parametrized in
estimation/discrimination problems. \textbf{(a)}~We show the last transmission
$a_{n}\rightarrow b_{n}$ through $\mathcal{E}$, which occurs between two
adaptive QOs $\Lambda_{n-1}$ and $\Lambda_{n}$. This last step produces the
output state $\rho_{n}$. \textbf{(b)}~In each transmission, we replace
$\mathcal{E}$ with its $M$-port simulation $\mathcal{E}^{M}$ so that the
output of the protocol becomes $\rho_{n}^{M}$ which approximates $\rho_{n}$
for large $M$. Note that, in the last transmission, the register state
$\rho_{\mathbf{ab}a_{n}}$ undergoes the transformation $\rho_{\mathbf{ab}%
b_{n}}=\mathcal{I}_{\mathbf{ab}}\otimes\mathcal{E}^{M}(\rho_{\mathbf{ab}a_{n}%
})$. \textbf{(c)}~Each propagation through $\mathcal{E}^{M}$ is replaced by
its PBT\ simulation. For the last transmission, this means that $\rho
_{\mathbf{ab}b_{n}}=\mathcal{I}_{\mathbf{ab}}\otimes\mathcal{T}^{M}%
(\rho_{\mathbf{ab}a_{n}}\otimes\rho_{\mathcal{E}}^{\otimes M})$ where
$\mathcal{T}^{M}$ is the LOCC of the PBT and $\rho_{\mathcal{E}}$ is the Choi
matrix of the original channel. \textbf{(d)}~All the adaptive QOs $\Lambda
_{i}$ and the simulation LOCCs $\mathcal{T}^{M}$ are collapsed into a single
(trace-preserving) QO $\bar{\Lambda}$. Correspondingly, $n$ instances of
$\rho_{\mathcal{E}}^{\otimes M}$ are collected. As a result, the approximate
output $\rho_{n}^{M}$ is given by $\bar{\Lambda}$ applied to the
tensor-product state $\rho_{\mathcal{E}}^{\otimes nM}$ as in Eq.~(\ref{sss}).}%
\label{stretch}%
\end{figure*}


\subsection{PBT stretching of an adaptive protocol}

Channel simulation is a preliminary tool for the following technique of
teleportation stretching, where an arbitrary adaptive protocol is reduced into
a simpler block version. There are two main steps. First of all, we need to
replace each channel $\mathcal{E}$ with its $M$-port approximation
$\mathcal{E}^{M}$\ while controlling the propagation of the simulation error
$\delta_{M}$ from the channel to the output state. This step is crucial also
in simulations via standard teleportation~\cite{TQC,ReviewMETRO} (see also
Refs.~\cite{networkPIRS,Multipoint,finiteStretching,nonPauli,HWchannels}).
Second, we need to \textquotedblleft stretch\textquotedblright\ the
protocol~\cite{PLOB} by replacing the various instances of the approximate
channel $\mathcal{E}^{M}$ with a collection of Choi matrices $\rho
_{\mathcal{E}}^{\otimes M}$ and then suitably re-organizing all the remaining
QOs. Here we describe the technique for a generic task, before specifying it
to QCD.

Given an adaptive protocol $\mathcal{P}_{n}$ over a channel $\mathcal{E}$ with
output $\rho_{n}$, consider the same protocol over the simulated channel
$\mathcal{E}^{M}$, so that we get the different output $\rho_{n}^{M}$. Using a
\textquotedblleft peeling\textquotedblright\ argument (see Methods), we bound
the output error in terms of the channel simulation error
\begin{equation}
||\rho_{n}-\rho_{n}^{M}||\leq n||\mathcal{E}-\mathcal{E}^{M}||_{\diamond}\leq
n\delta_{M}. \label{outputERROR}%
\end{equation}
Once understood that the output state can be closely approximated, let us
simplify the adaptive protocol over $\mathcal{E}^{M}$. Using the simulation in
Eq.~(\ref{LOCCsim}), we may replace each channel $\mathcal{E}^{M}$ with the
resource state $\rho_{\mathcal{E}}^{\otimes M}$, iterate the process for all
$n$ uses, and collapse all the simulation LOCCs and QOs as shown in
Fig.~\ref{stretch}. As a result, we may write the multi-copy Choi
decomposition\
\begin{equation}
\rho_{n}^{M}=\bar{\Lambda}(\rho_{\mathcal{E}}^{\otimes nM})~, \label{sss}%
\end{equation}
for a trace-preserving QO $\bar{\Lambda}$. Now, we can combine the two
ingredients of Eqs.~(\ref{outputERROR}) and~(\ref{sss}), into the following.

\begin{lemma}
[PBT stretching]\label{lemma}Consider an adaptive quantum protocol (with
arbitrary task) over an arbitrary $d$-dimensional quantum channel
$\mathcal{E}$ (which may be unknown and parametrized). After $n$ uses, the
output $\rho_{n}$ of the protocol can be decomposed as follows
\begin{equation}
||\rho_{n}-\bar{\Lambda}(\rho_{\mathcal{E}}^{\otimes nM})||\leq n\delta_{M},
\label{LemmaEQ}%
\end{equation}
where $\bar{\Lambda}$ is a trace-preserving QO, $\rho_{\mathcal{E}}$ is the
Choi matrix of $\mathcal{E}$, and $\delta_{M}$ is the $M$-port simulation
error in Eq.~(\ref{kbound}).
\end{lemma}

When we apply the lemma to protocols of quantum or private communication,
where the QOs $\Lambda_{i}$ are LOCCs, then we may write Eq.~(\ref{LemmaEQ})
with $\bar{\Lambda}$ being a LOCC. In protocols of channel estimation or
discrimination, where $\mathcal{E}$ is parametrized, we may write
Eq.~(\ref{LemmaEQ}) with $\rho_{\mathcal{E}}$ storing the parameter of the
channel. In particular, for QCD we have $\{\mathcal{E}_{u}\}_{u=0,1}$ and the
output $\rho_{n}(u)$ of the adaptive protocol $\mathcal{P}_{n}$ can be
decomposed as follows
\begin{equation}
||\rho_{n}(u)-\bar{\Lambda}(\rho_{\mathcal{E}_{u}}^{\otimes nM})||\leq
n\delta_{M}.
\end{equation}

\subsection{Ultimate bound for channel discrimination}

We are now ready to show the lower bound for minimum error probability
$p_{n}(\mathcal{E}_{0}\neq\mathcal{E}_{1})$ in Eq.~(\ref{eqDD}). Consider an
arbitrary protocol $\mathcal{P}_{n}$, for which we may write
Eq.~(\ref{protPROB}). Combining Lemma~2 with the triangle inequality leads to%
\begin{align}
||\rho_{n}(0)-\rho_{n}(1)||  &  \leq2n\delta_{M}+||\bar{\Lambda}%
(\rho_{\mathcal{E}_{0}}^{\otimes nM})-\bar{\Lambda}(\rho_{\mathcal{E}_{1}%
}^{\otimes nM})||\nonumber\\
&  \leq2n\delta_{M}+||\rho_{\mathcal{E}_{0}}^{\otimes nM}-\rho_{\mathcal{E}%
_{1}}^{\otimes nM}||, \label{eqBOUND}%
\end{align}
where we also use the monotonicity of the trace distance under channels.
Because $\bar{\Lambda}$\ is lost, the bound does no longer depend on the
details of the protocol $\mathcal{P}_{n}$, which means that it applies to all
adaptive protocols. Thus, using Eq.~(\ref{eqBOUND}) in Eqs.~(\ref{protPROB})
and~(\ref{protINF}), we get the following.

\begin{theorem}
\label{mainTHEO}Consider the adaptive discrimination of two channels
$\{\mathcal{E}_{u}\}_{u=0,1}$ in dimension $d$. After $n$ probings, the
minimum error probability satisfies the bound%
\begin{equation}
p_{n}(\mathcal{E}_{0}\neq\mathcal{E}_{1})\geq B:=\frac{1-n\delta_{M}%
-D(\rho_{\mathcal{E}_{0}}^{\otimes nM},\rho_{\mathcal{E}_{1}}^{\otimes nM}%
)}{2}, \label{surprise}%
\end{equation}
where $M$ may be chosen to maximize the right hand side.
\end{theorem}

\noindent Not only this is the first universal bound for adaptive QCD, but
also its analytical form is rather surprising. In fact, its tighest value is
given by an optimal (finite) number of ports $M$ for the underlying protocol
of PBT.

Let us bound the trace distance in Eq.~(\ref{surprise}) as
\begin{equation}
D^{2}\leq1-F^{2nM},~F:=\mathrm{Tr}\sqrt{\sqrt{\rho_{\mathcal{E}_{0}}}%
\rho_{\mathcal{E}_{1}}\sqrt{\rho_{\mathcal{E}_{0}}}}, \label{letus}%
\end{equation}
where $F$ is the fidelity between the Choi matrices of the
channels. This comes from the Fuchs-van de Graaf
relations~\cite{Fuchs} and the multiplicativity of the fidelity
over tensor products. Other bounds that can be written are
\begin{equation}
D\leq nM\left\Vert
\rho_{\mathcal{E}_{0}}-\rho_{\mathcal{E}_{1}}\right\Vert ,
\end{equation}
from the subadditivity of the trace distance, and
\begin{equation}
D\leq\sqrt{nM(\ln\sqrt{2})\min\{S(\rho_{\mathcal{E}_{0}}||\rho_{\mathcal{E}_{1}}),S(\rho_{\mathcal{E}_{1}}||\rho_{\mathcal{E}_{0}})\}},
\end{equation}
from the Pinsker inequality~\cite{Pin1,Pin2}, where $S(\rho
||\sigma)=\mathrm{Tr}[\rho(\log_{2}\rho-\log_{2}\sigma)]$ is the
relative entropy~\cite{NiCh}.

If we exploit Eqs.~(\ref{kbound}) and~(\ref{letus}) in Eq.~(\ref{surprise}),
we may write the following simplified bound%
\begin{equation}
B\geq\frac{1}{2}-\frac{\sqrt{1-F^{2nM}}}{2}-\frac{d(d-1)n}{M}\,.
\end{equation}
In the previous formula there are terms with opposite monotonicity in $M$, so
that the maximum value of the bound $B$ is achieved at some intermediate value
of $M$. Setting $M=xd(d-1)n$ for some $x>2$, we get
\begin{equation}
B\geq\frac{1}{2}-\frac{1}{x}-\frac{1}{2}\sqrt{1-F^{2xd(d-1)n^{2}}}.
\end{equation}
One good choice is therefore $M=4d(d-1)n$, so that
\begin{equation}
B\geq(1-2\sqrt{1-F^{8d(d-1)n^{2}}})/4.
\end{equation}

In particular, consider two infinitesimally-close channels, so that
$F\simeq1-\epsilon$ where $\epsilon\simeq0$ is the infidelity. By expanding in
$\epsilon$ for any finite $n$, we may write
\begin{equation}
B\geq\frac{1}{4}-n\sqrt{2d(d-1)\epsilon}\simeq\frac{\exp(-4n\sqrt
{2d(d-1)\epsilon})}{4}. \label{boundB}%
\end{equation}
For instance, in the case of qubits this becomes $[\exp
(-8n\sqrt{\epsilon})]/4$, to be compared with the upper bound
$[\exp(-2n\epsilon)]/2$ computed from Eq.~(\ref{fidSIM}).
Discriminating between two close quantum channels is a problem in
many physical scenarios. For instance, this is typical in quantum
optical resolution~\cite{Tsang15,Lupo16,Tsang2} (discussed below),
quantum
illumination~\cite{Qill0,Qill1,Qill2,Qill3,Qill4,Qill5,Qill6,Qill7,dd3,Qill8}
(discussed below), ideal quantum
reading~\cite{Qread,QreadCAP,Arno12,ArnoIJQI,GaeENTROPY}, quantum
metrology~\cite{Sam1,Sam2,Paris,Giova,ReviewNEW} (discussed
below), and also tests of quantum field theories in non-inertial
frames~\cite{Doukas}, e.g., for detecting effects such as the
Unruh or the Hawking radiation.

\subsection{Limits of single-photon quantum optical resolution}

Consider a microscope-type problem where we aim at locating a
point in two possible positions, either $s/2$ or $-s/2$, where the
separation $s$ is very small. Assume we are limited to use probe
states with at most one photon and an output finite-aperture
optical system (this makes the optical process to be a
qubit-to-qutrit channel, so that the input dimension is $d=2$).
Apart from this, we are allowed to use an arbitrary large quantum
computer and arbitrary QOs to manipulate its registers. We may
apply Eq.~(\ref{boundB}) with $\epsilon\simeq\eta s^{2}/16$, where
$\eta$ is a diffraction-related loss parameter. In this way, we
find that the error probability affecting the discrimination of
the two positions is approximately bounded by $B\gtrsim
\frac{1}{4}\exp(-2ns\sqrt{\eta})$. This bound establishes a no-go
for perfect quantum optical resolution. See Supplementary
Section~II for more mathematical details on this specific
application.

\subsection{Limits of adaptive quantum illumination}

Consider the protocol of quantum illumination in the DV setting~\cite{Qill0}.
Here the problem is to discriminate the presence or not of a target with low
reflectivity $\eta\simeq0$ in a thermal background which has $b\ll1$ mean
thermal photons per optical mode. One assumes that $d$ modes are used in each
probing of the target and each of them contains at most one photon. This means
that the Hilbert space is $(d+1)$-dimensional with basis $\{\left\vert
0\right\rangle ,\left\vert 1\right\rangle ,\ldots,\left\vert d\right\rangle
\}$, where $\left\vert i\right\rangle :=\left\vert 0\cdots010\cdots
0\right\rangle $ has one photon in the $i$th mode. If the target is absent
($u=0$), the receiver detects thermal noise; if the target is present ($u=1$),
the receiver measures a mixture of signal and thermal noise.

In the most general (adaptive) version of the protocol, the
receiver belongs to a large quantum computer where the
$(d+1)$-dimensional signal qudits are picked from an input
register, sent to target, and their reflection stored in an output
register, with adaptive QOs performed between each probing. After
$n$ probings, the state of the registers $\rho_{n}(u)$ is
optimally detected. Assuming the typical regime of quantum
illumination~\cite{Qill0}, we find that the error probability
affecting target detection is approximately bounded by
$B\gtrsim\frac{1}{4}\exp(-4nd\sqrt{\eta})$. This bound establishes
a no-go for exponential improvement in quantum illumination.
Entanglement and adaptiveness can \textit{at most} improve the
error exponent with respect to separable probes, for which the
error probability is $\lesssim\frac{1}{2}\exp [-n\eta/(8d)]$. See
also Supplementary Section~III.

\subsection{Limits of adaptive quantum metrology}

Consider the adaptive estimation of a continuous parameter $\theta$ encoded in
a quantum channel $\mathcal{E}_{\theta}$. After $n$ probings, we have a
$\theta$-dependent output state $\rho_{n}(\theta)$ generated by an adaptive
quantum estimation protocol $\mathcal{P}_{n}$. This output state is then
measured by a POVM $\mathcal{M}$ providing an optimal unbiased estimator
$\tilde{\theta}$ of parameter $\theta$. The minimum error variance
Var$(\tilde{\theta}):=\langle(\tilde{\theta}-\theta)^{2}\rangle$ must satisfy
the quantum Cramer-Rao bound \textrm{Var}$(\tilde{\theta})\geq1/$%
\textrm{QFI}$_{\theta}(\mathcal{P}_{n})$, where \textrm{QFI}$_{\theta
}(\mathcal{P}_{n})$ is the quantum Fisher information~\cite{Sam1} associated
with $\mathcal{P}_{n}$. The ultimate precision of adaptive quantum metrology
is given by the optimization over all protocols
\begin{equation}
\overline{\text{\textrm{QFI}}}_{\theta}^{n}:=\sup_{\mathcal{P}_{n}%
}\text{\textrm{QFI}}_{\theta}(\mathcal{P}_{n}). \label{fisheroptm}%
\end{equation}

This quantity can be simplified by PBT\ stretching. In fact, for any input
state $\rho_{C}$, we may write the simulation $\mathcal{E}_{\theta}^{M}%
(\rho_{C})=\mathcal{T}^{M}(\rho_{C}\otimes\rho_{\mathcal{E}_{\theta}}^{\otimes
M})$ which is an immediate extension of Eq.~(\ref{LOCCsim}). In this way, the
output state can be decomposed following Lemma~2, i.e., we may write
$||\rho_{n}(\theta)-\bar{\Lambda}(\rho_{\mathcal{E}_{\theta}}^{\otimes
nM})||\leq n\delta_{M}$. Exploiting the latter inequality for large $n$, we
find that the ultimate bound of adaptive quantum metrology takes the form
\begin{equation}
\overline{\text{\textrm{QFI}}}_{\theta}^{n}\lesssim n^{2}\mathrm{QFI}%
(\rho_{\mathcal{E}_{\theta}}), \label{cubem}%
\end{equation}
where $\mathrm{QFI}(\rho_{\mathcal{E}_{\theta}})$ is computed on
the channel's Choi matrix. In particular, we see that PBT allows
us to write a simple bound in terms of the Choi matrix and implies
a general no-go theorem for super-Heisenberg scaling in quantum
metrology. See Supplementary Section~IV for a detailed proof of
Eq.~(\ref{cubem}).

\subsection{Tightening the main formula}

Let us note that the formula in Theorem~\ref{mainTHEO} is expressed in terms
of the universal error $\delta_{M}$ coming from the PBT simulation of the
identity channel (Lemma~\ref{lemmaPBT}). There are situations where the
diamond distance $\Delta_{M}:=||\mathcal{E}-\mathcal{E}^{M}||_{\diamond}$
between a quantum channel $\mathcal{E}$ and its $M$-port simulation
$\mathcal{E}^{M}$ is exactly computable. In these cases, we can certainly
formulate a tighter version of Eq.~(\ref{surprise}) where $\delta_{M}$ is
suitably replaced. In fact, from the peeling argument, we have $||\rho
_{n}-\rho_{n}^{M}||\leq n\Delta_{M}$, so that a tighter version of
Eq.~(\ref{LemmaEQ}) is simply $||\rho_{n}-\bar{\Lambda}(\rho_{\mathcal{E}%
}^{\otimes nM})||\leq n\Delta_{M}$. Then, for the two possible outputs
$\rho_{n}(0)$ and $\rho_{n}(1)$ of an adaptive discrimination protocol over
$\mathcal{E}_{0}$ and $\mathcal{E}_{1}$, we can replace Eq.~(\ref{eqBOUND})
with
\begin{equation}
||\rho_{n}(0)-\rho_{n}(1)||\leq2n\bar{\Delta}_{M}+||\rho_{\mathcal{E}_{0}%
}^{\otimes nM}-\rho_{\mathcal{E}_{1}}^{\otimes nM}||,
\end{equation}
where $\bar{\Delta}_{M}:=(||\mathcal{E}_{0}-\mathcal{E}_{0}^{M}||_{\diamond
}+||\mathcal{E}_{1}-\mathcal{E}_{1}^{M}||_{\diamond})/2$. It is now easy to
check that Eq.~(\ref{surprise}) becomes the following%
\begin{equation}
p_{n}(\mathcal{E}_{0}\neq\mathcal{E}_{1})\geq\frac{1-n\bar{\Delta}_{M}%
-D(\rho_{\mathcal{E}_{0}}^{\otimes nM},\rho_{\mathcal{E}_{1}}^{\otimes nM}%
)}{2}. \label{improvedB}%
\end{equation}
In the following section, we show that $\bar{\Delta}_{M}$, and therefore the
bound in Eq.~(\ref{improvedB}), can be computed for the discrimination of
amplitude damping channels.

\subsection{Discrimination of amplitude damping channels}

As an additional example of application of the bound, consider the
discrimination between amplitude damping channels. These channels are not
teleportation covariant, so that the results from Ref.~\cite{PirCo} do not
apply and no bound is known on the error probability for their adaptive
discrimination. Recall that an amplitude damping channel $\mathcal{E}_{p}$
transforms an input state $\rho$ as follows%
\begin{equation}
\mathcal{E}_{p}(\rho)=%
{\textstyle\sum\nolimits_{i=0,1}}
K_{i}\rho K_{i}^{\dagger},
\end{equation}
with Kraus operators%
\begin{equation}
K_{0}:=\left\vert 0\right\rangle \left\langle 0\right\vert +\sqrt
{1-p}\left\vert 1\right\rangle \left\langle 1\right\vert ,~K_{1}:=\sqrt
{p}\left\vert 0\right\rangle \left\langle 1\right\vert ,
\end{equation}
where $\{\left\vert 0\right\rangle ,\left\vert 1\right\rangle \}$ is the
computational basis and $p$ is the damping probability or rate.

Given two amplitude damping channels, $\mathcal{E}_{p_{0}}$ and $\mathcal{E}%
_{p_{1}}$, first assume a discrimination protocol where these
channels are probed by $n$ maximally-entangled states and the
outputs are optimally measured. The optimal error probability for
this (non-adaptive) block protocol is given by
$p_{n}^{\text{block}}=[1-D(\rho_{\mathcal{E}_{p_{0}}}^{\otimes
n},\rho_{\mathcal{E}_{p_{1}}}^{\otimes n})]/2$ and satisfies
\begin{equation}
\frac{1-\sqrt{1-F(p_{0},p_{1})^{2n}}}{2}\leq p_{n}^{\text{block}}\leq
\frac{F(p_{0},p_{1})^{n}}{2}, \label{iidPROB}%
\end{equation}
where $F(p_{0},p_{1}):=F(\rho_{\mathcal{E}_{p_{0}}},\rho_{\mathcal{E}_{p_{1}}%
})$ is the fidelity between the Choi matrices. In particular, we explicitly
compute%
\begin{equation}
F=\frac{1+\sqrt{(1-p_{0})(1-p_{1})}+\sqrt{p_{0}p_{1}}}{2}.
\end{equation}
It is clear that $p_{n}^{\text{block}}$ in Eq.~(\ref{iidPROB}) is an upper
bound to ultimate (adaptive) error probability $p_{n}(\mathcal{E}_{p_{0}}%
\neq\mathcal{E}_{p_{1}})$ for the discrimination of the two channels.

To lowerbound the ultimate probability we employ
Eq.~(\ref{improvedB}). In fact, for the $M$-port simulation
$\mathcal{E}_{p}^{M}$ of $\mathcal{E}_{p}$, we
compute%
\begin{equation}
\Delta_{M}(p)=||\mathcal{E}_{p}-\mathcal{E}_{p}^{M}||_{\diamond} =\xi
_{M}\left( \frac{1-p}{2}+\sqrt{1-p}\right) ,\label{Deltadamp}%
\end{equation}
where $\xi_{M}$ are the PBT numbers defined in Eq.~(\ref{PBTnumbers}). For any
two amplitude damping channels, $\mathcal{E}_{p_{0}}$ and $\mathcal{E}_{p_{1}%
}$, we can then compute $\bar{\Delta}_{M}(p_{0},p_{1})$ and use
Eq.~(\ref{improvedB}) to bound $p_{n}(\mathcal{E}_{p_{0}}\neq\mathcal{E}%
_{p_{1}})$. More precisely, we can also exploit Eq.~(\ref{letus}) and write
the computable lower bound
\begin{equation}
p_{n}(\mathcal{E}_{p_{0}}\neq\mathcal{E}_{p_{1}})\geq\frac{1-n\bar{\Delta}%
_{M}(p_{0},p_{1})-\sqrt{1-F(p_{0},p_{1})^{2nM}}}{2}. \label{LBcompute}%
\end{equation}

\begin{figure}[ptb]
\begin{center}
\vspace{+0.2cm} \includegraphics[width=0.45\textwidth]{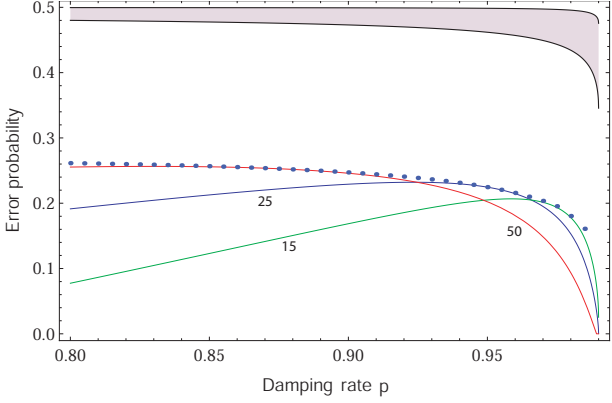}
\vspace{-0.1cm}
\end{center}
\caption{Error probability in the discrimination of two amplitude
damping channels, one with damping rate $p\geq0.8$ and the other
with rate $p+1\%$. We assume $n=20$ probings of the unknown
channel. The upper dark region identifies the region where the
error probability $p_{n}^{\text{block}}$ of
Eq.~(\ref{iidPROB}) lies. The adaptive error probability $p_{n}(\mathcal{E}%
_{p_{0}}\neq\mathcal{E}_{p_{1}})$ lies below this dark region and
above the dotted points, which represent our lower bound of
Eq.~(\ref{LBcompute}) optimized over the number of ports $M$. For
comparison, we also plot the lower bound for specific $M$.}
\label{Figcomp}%
\end{figure}

In Fig.~\ref{Figcomp} we show an example of discrimination between two
amplitude damping channels. In particular, we show how large is the gap
between the upper bound $p_{n}^{\text{block}}$ of Eq.~(\ref{iidPROB}) and the
lower bound in Eq.~(\ref{LBcompute}) suitably optimized over the number of
ports $M$. It is an open question to find exactly $p_{n}(\mathcal{E}_{p_{0}%
}\neq\mathcal{E}_{p_{1}})$. At this stage, we do not know if this result may
achieved by tightening the upper bound or the lower bound.

\section{Discussion}

In this work we have established a general and fundamental lower
bound for the error probability affecting the adaptive
discrimination of two arbitrary quantum channels acting on a
finite-dimensional Hilbert space. This bound is conveniently
expressed in terms of the Choi matrices of the channels involved,
so that it is very easy to compute. It also applies to many
scenarios, including adaptive protocols for quantum-enhance
optical resolution and quantum illumination. In order to derive
our result, we have employed port-based teleportation as a tool
for channel simulation, and developed a methodology which
simplifies adaptive protocols performed over an arbitrary
finite-dimensional channel. This technique can be applied to many
other scenarios. For instance, in quantum metrology we are able to
prove that adaptive protocols of quantum channel estimation are
limited by a bound simply expressed in terms of the Choi matrix of
the channel and following the Heisenberg scaling in the number of
probings. Not only this shows that our bound is asymptotically
tight but also draws an unexpected connection between port-based
teleportation and quantum metrology. Further potential
applications are in quantum and private communications, which are
briefly discussed in our Supplementary Section~V.

\section{Methods}

\subsection{Simulation error in diamond norm (proof of Lemma~1)\label{SUP0}}

It is easy to check that the channel $\Gamma_{M}$ associated with the qudit
PBT protocol of Ref.~\cite{PBT} is covariant under unitary transformations,
i.e.,
\begin{equation}
\Gamma_{M}(U\rho U^{\dag})=U\Gamma_{M}(\rho)U^{\dag},
\end{equation}
for any input state $\rho$ and unitary operator $U$. As discussed
in Ref.~\cite{Majenz}, for a channel with such a symmetry, the
diamond distance with the identity map is saturated by a maximally
entangled state, i.e.,
\begin{equation}
\Vert\mathcal{I}-\Gamma_{M}\Vert_{\diamond}=\Vert|\Phi\rangle\langle
\Phi|-\mathcal{I}\otimes\Gamma_{M}\left(  |\Phi\rangle\langle\Phi|\right)
\Vert\,, \label{CMajenz}%
\end{equation}
where $|\Phi\rangle=d^{-1/2}\sum_{k=1}^{d}|k\rangle|k\rangle$. Here we first
show that
\begin{equation}
\Vert|\Phi\rangle\langle\Phi|-\mathcal{I}\otimes\Gamma_{M}\left(  |\Phi
\rangle\langle\Phi|\right)  \Vert=2[1-f_{e}(\Gamma_{M})]~. \label{showCL}%
\end{equation}

In fact, note that the map $\Lambda_{M}=\mathcal{I}\otimes\Gamma_{M}$ is
covariant under twirling unitaries of the form $U\otimes U^{\ast}$, i.e.,
\begin{align}
&  \Lambda_{M}\left[  (U\otimes U^{\ast})\rho(U\otimes U^{\ast})^{\dag
}\right]  \nonumber\\
&  =(U\otimes U^{\ast})\Lambda_{M}(\rho)(U\otimes U^{\ast})^{\dag},
\end{align}
for any input state $\rho$ and unitary operator $U$. This implies that the
state $\Lambda_{M}(|\Phi\rangle\langle\Phi|)$ is invariant under twirling
unitaries, i.e.,
\begin{equation}
(U\otimes U^{\ast})\Lambda_{M}(|\Phi\rangle\langle\Phi|)(U\otimes U^{\ast
})^{\dag}=\Lambda_{M}(|\Phi\rangle\langle\Phi|)~.
\end{equation}
This is therefore an isotropic state of the form%
\begin{equation}
\Lambda_{M}(|\Phi\rangle\langle\Phi|)=(1-p)|\Phi\rangle\langle\Phi|+\frac
{p}{d^{2}}\mathbb{I},\label{eq:ADchoi1}%
\end{equation}
where $\mathbb{I}$ is the two-qudit identity operator.

We may rewrite this state as follows%
\begin{equation}
\Lambda_{M}(|\Phi\rangle\langle\Phi|)=F|\Phi\rangle\langle\Phi|+(1-F)\rho
^{\perp},\label{deco1}%
\end{equation}
where $\rho^{\perp}$ is state with support in the orthogonal complement of
$\Phi$, and $F$ is the singlet fraction%
\begin{equation}
F:=\langle\Phi|\Lambda_{M}(|\Phi\rangle\langle\Phi|)|\Phi\rangle=1-p+pd^{-2}.
\end{equation}
Thanks to the decomposition in Eq.~(\ref{deco1}) and using basic properties of
the trace norm~\cite{NiCh}, we may then write
\begin{align}
&  \Vert|\Phi\rangle\langle\Phi|-\Lambda_{M}\left(  |\Phi\rangle\langle
\Phi|\right)  \Vert\nonumber\\
&  =\Vert(1-F)|\Phi\rangle\langle\Phi|-(1-F)\rho^{\perp}\Vert\nonumber\\
&  =(1-F)\Vert|\Phi\rangle\langle\Phi|\Vert+(1-F)\Vert\rho^{\perp}%
\Vert\nonumber\\
&  =2(1-F)\nonumber\\
&  =2[1-f_{e}(\Gamma_{M})],
\end{align}
where the last step exploits the fact that the singlet fraction $F$ is the
channel's entanglement fidelity $f_{e}(\Gamma_{M})$. This completes the proof
of Eq.~(\ref{showCL}).

Therefore, combining Eqs.~(\ref{CMajenz}) and (\ref{showCL}), we obtain
\begin{equation}
\Vert\mathcal{I}-\Gamma_{M}\Vert_{\diamond}=2[1-f_{e}(\Gamma_{M}%
)],\label{tttt}%
\end{equation}
which is Eq.~(\ref{firstEQQ}) of the main text. Then, we know that the
entanglement fidelity of $\Gamma_{M}$ is bounded as~\cite{PBT}
\begin{equation}
f_{e}(\Gamma_{M})\geq1-d(d-1)M^{-1}.\label{Hiro}%
\end{equation}
Therefore, using Eq.~(\ref{Hiro}) in Eq.~(\ref{tttt}), we derive the following
upper bound
\begin{equation}
\Vert\mathcal{I}-\Gamma_{M}\Vert_{\diamond}\leq2d(d-1)M^{-1},
\end{equation}
which is Eq.~(\ref{kbound}) of the main text.

Let us now prove Eq.~(\ref{exactBB}). It is known~\cite{PBT1} that
implementing the standard PBT\ protocol over the resource state of
Eq.~(\ref{nonOPT}) leads to a PBT channel $\Gamma_{M}$ which is a
qudit depolarizing channel. Its isotropic Choi matrix
$\rho_{\Gamma_{M}}$, given in Eq.~(\ref{eq:ADchoi1}), can be
written in the form
\begin{equation}
\rho_{\Gamma_{M}}=\left(  1-\frac{d^{2}-1}{d^{2}}\xi_{M}\right)  |\Phi
\rangle^{0}\langle\Phi|+\sum_{i=1}^{d^{2}-1}\frac{\xi_{M}}{d^{2}}|\Phi
\rangle^{i}\langle\Phi|,
\end{equation}
where $\xi_{M}$ is the probability $p$ of depolarizing, $|\Phi\rangle
^{0}\langle\Phi|$ is the projector onto the initial maximally-entangled state
of two qudits (one system of which was sent through the channel), and
$|\Phi\rangle^{i}\langle\Phi|$ are the projectors onto the other $d^{2}-1$
maximally-entangled states of two qudits (generalized Bell states). Since the
Choi matrix of the identity channel is $\rho_{\mathcal{I}}=|\Phi\rangle
^{0}\langle\Phi|$, it is easy to compute
\begin{align}
\left\vert \rho_{\mathcal{I}}-\rho_{\Gamma_{M}}\right\vert  &  :=\sqrt{\left(
\rho_{\mathcal{I}}-\rho_{\Gamma_{M}}\right)  \left(  \rho_{\mathcal{I}}%
-\rho_{\Gamma_{M}}\right)  ^{\dag}}\nonumber\\
&  =\frac{d^{2}-1}{d^{2}}\xi_{M}|\Phi\rangle^{0}\langle\Phi|+\sum_{i=1}%
^{d^{2}-1}\frac{\xi_{M}}{d^{2}}|\Phi\rangle^{i}\langle\Phi|.
\end{align}

From the previous equation, we derive
\begin{equation}
\mathrm{Tr}_{2}\left\vert \rho_{\mathcal{I}}-\rho_{\Gamma_{M}}\right\vert
=\frac{2\left(  d^{2}-1\right)  }{d^{3}}\xi_{M}\sum_{j=0}^{d-1}|j\rangle
\langle j|,\label{null}%
\end{equation}
where we have used $\mathrm{Tr}_{2}|\Phi\rangle^{i}\langle\Phi|=d^{-1}%
\sum_{j=0}^{d-1}|j\rangle\langle j|$ in the qudit computational
basis $\{|j\rangle\}$ and we have summed over the $d^{2}$
generalized Bell states. It is clear that Eq.~(\ref{null}) is a
diagonal matrix with equal non-zero elements, i.e., it is a
scalar. As a result, we can apply Proposition~1 of
Ref.~\cite{nechita} over the Hermitian operator $\rho_{\mathcal{I}}%
-\rho_{\Gamma_{M}}$, and write
\begin{gather}
\Vert\mathcal{I}-\Gamma_{M}\Vert_{\diamond}=\Vert\rho_{\mathcal{I}}-\rho_{\Gamma_{M}}\Vert\nonumber\\
=\mathrm{Tr}\left\vert \rho_{\mathcal{I}}-\rho_{\Gamma_{M}}\right\vert
=\frac{2\left(  d^{2}-1\right)  }{d^{2}}\xi_{M}~.
\end{gather}

The final step of the proof is to compute the explicit expression
of $\xi_{M}$ for qubits, which is the formula given in
Eq.~(\ref{PBTnumbers}). Because this derivation is technically
involved, it is reported in Supplementary Section~I.

\subsection{Propagation of the simulation error\label{SUP3}}

For the sake of completeness, we provide the proof of the first
inequality in~Eq.~(\ref{outputERROR}) (this kind of proof already
appeared in Refs.~\cite{PLOB,TQC}). Consider the adaptive protocol
described in the main
text. For the $n$-use output state we may compactly write%
\begin{equation}
\rho_{n}=\Lambda_{n}\circ\mathcal{E}\circ\Lambda_{n-1}\circ\cdots
\circ\mathcal{E}\circ\Lambda_{1}\circ\mathcal{E}(\rho_{0}),
\end{equation}
where $\Lambda$'s are adaptive QOs and $\mathcal{E}$ is the channel applied to
the transmitted signal system. Then, $\rho_{0}$ is the preparation state of
the registers, obtained by applying the first\ QO$\ \Lambda_{0}$ to some
fundamental state. Similarly, for the $M$-port simulation of the protocol, we
may write%
\begin{equation}
\rho_{n}^{M}=\Lambda_{n}\circ\mathcal{E}^{M}\circ\Lambda_{n-1}\circ\cdots
\circ\mathcal{E}^{M}\circ\Lambda_{1}\circ\mathcal{E}^{M}(\rho_{0}),
\end{equation}
where $\mathcal{E}^{M}$ is in the place of $\mathcal{E}$. 

Consider now two instances ($n=2$) of the adaptive protocol. We may bound the
trace distance between $\rho_{2}$ and $\rho_{2}^{M}$ using a \textquotedblleft
peeling\textquotedblright\ argument~\cite{PirCo,PLOB,TQC,Uniform,ReviewMETRO}
\begin{align}
\left\Vert \rho_{2}-\rho_{2}^{M}\right\Vert  &  =\left\Vert \Lambda_{2}%
\circ\mathcal{E}\circ\Lambda_{1}\circ\mathcal{E}(\rho_{0})\right. \nonumber\\
&  -\Lambda_{2}\circ\mathcal{E}^{M}\circ\Lambda_{1}\circ\mathcal{E}^{M}%
(\rho_{0})||\nonumber\\
&  \overset{{\tiny (1)}}{\leq}||\mathcal{E}\circ\Lambda_{1}\circ
\mathcal{E}(\rho_{0})-\mathcal{E}^{M}\circ\Lambda_{1}\circ\mathcal{E}^{M}%
(\rho_{0})||\nonumber\\
&  \overset{{\tiny (2)}}{\leq}||\mathcal{E}\circ\Lambda_{1}\circ
\mathcal{E}(\rho_{0})-\mathcal{E}\circ\Lambda_{1}\circ\mathcal{E}^{M}(\rho
_{0})||\nonumber\\
&  +||\mathcal{E}^{M}\circ\Lambda_{1}\circ\mathcal{E}(\rho_{0})-\mathcal{E}%
^{M}\circ\Lambda_{1}\circ\mathcal{E}^{M}(\rho_{0})||\nonumber\\
&  \overset{{\tiny (3)}}{\leq}||\mathcal{E}(\rho_{0})-\mathcal{E}^{M}(\rho
_{0})||\nonumber\\
&  +||\mathcal{E}[\Lambda_{1}\circ\mathcal{E}^{M}(\rho_{0})]-\mathcal{E}%
^{M}[\Lambda_{1}\circ\mathcal{E}^{M}(\rho_{0})]||\nonumber\\
&  \overset{{\tiny (4)}}{\leq}2||\mathcal{E}-\mathcal{E}^{M}||_{\diamond}~.
\label{DiamondV}%
\end{align}
In $(1)$ we use the monotonicity of the trace distance under
completely-positive trace-preserving (CPTP) maps (i.e., quantum
channels); in $(2)$ we employ the triangle inequality; in $(3)$ we
use the monotonicity with respect to the the CPTP map
$\mathcal{E}\circ\Lambda_{1}$ whereas in $(4)$ we exploit the fact
that the diamond norm is an upper bound for the trace norm
computed on any input state. Generalizing the result of
Eq.~(\ref{DiamondV}) to arbitrary $n$, we achieve the first
inequality in Eq.~(\ref{outputERROR}). Note that the previous
reasoning also applies to a classically-parametrized channel
$\mathcal{E}_{u}$.

\subsection{PBT simulation of amplitude damping channels}

Here we show the result in Eq.~(\ref{Deltadamp}) for $\Delta_{M}%
(p)=||\mathcal{E}_{p}-\mathcal{E}_{p}^{M}||_{\diamond}$, which is the error
associated with the $M$-port simulation of an arbitrary amplitude damping
channel $\mathcal{E}_{p}$. From Ref.~\cite{PBT1}, we know that the PBT channel
$\Gamma^{M}$ is a depolarizing channel. In the qubit computational basis
$\{\left\vert i,j\right\rangle \}_{i,j=0,1}$, it has the following Choi
matrix
\begin{equation}
\rho_{\Gamma^{M}}=%
\begin{pmatrix}
\frac{1}{2}-\frac{\xi_{M}}{4} & 0 & 0 & \frac{1}{2}-\frac{\xi_{M}}{2}\\
0 & \frac{\xi_{M}}{4} & 0 & 0\\
0 & 0 & \frac{\xi_{M}}{4} & 0\\
\frac{1}{2}-\frac{\xi_{M}}{2} & 0 & 0 & \frac{1}{2}-\frac{\xi_{M}}{4}%
\end{pmatrix}
,
\end{equation}
where $\xi_{M}$ are the PBT\ numbers of Eq.~(\ref{PBTnumbers}). Note that
these take decreasing positive values, for instance%
\begin{align}
\xi_{2} &  =\frac{6-\sqrt{3}}{6}\simeq0.71,\nonumber\\
\xi_{3} &  =1/2,\nonumber\\
\xi_{4} &  =\frac{13-2\sqrt{2}-2\sqrt{5}}{16},\nonumber\\
\xi_{5} &  =\frac{35-4\sqrt{6}-4\sqrt{10}}{48},\nonumber\\
\xi_{6} &  =\frac{70-15\sqrt{3}-5\sqrt{7}-3\sqrt{15}}{96}\simeq0.2.
\end{align}

By applying the Kraus operators $K_{0}$ and $K_{1}$ of $\mathcal{E}_{p}$
locally to $\rho_{\Gamma^{M}}$ we obtain the Choi matrix of the $M$-port
simulation $\mathcal{E}_{p}^{M}$, which is
\begin{equation}
\rho_{\mathcal{E}_{p}^{M}}=%
\begin{pmatrix}
x & 0 & 0 & y\\
0 & \left(  1-p\right)  \xi_{M} & 0 & 0\\
0 & 0 & w & 0\\
y & 0 & 0 & z
\end{pmatrix}
,
\end{equation}
where $x:=\frac{1}{2}-\left(  1-p\right)  \frac{\xi_{M}}{4}$, $y:=\sqrt
{1-p}\left(  \frac{1}{2}-\frac{\xi_{M}}{2}\right)  $, $z:=\left(  \frac{1}%
{2}-\frac{\xi_{M}}{4}\right)  \left(  1-p\right)  $, and $w:=\left(  \frac
{1}{2}-\frac{\xi_{M}}{4}\right)  p+\frac{\xi_{M}}{4}$. This has to be compared
with the Choi matrix of $\mathcal{E}_{p}$, which is%
\begin{equation}
\rho_{\mathcal{E}_{p}}=%
\begin{pmatrix}
\frac{1}{2} & 0 & 0 & \frac{\sqrt{1-p}}{2}\\
0 & 0 & 0 & 0\\
0 & 0 & \frac{p}{2} & 0\\
\frac{\sqrt{1-p}}{2} & 0 & 0 & \frac{1-p}{2}%
\end{pmatrix}
.
\end{equation}

Now, consider the Hermitian matrix $J=\rho_{\mathcal{E}_{p}^{M}}%
-\rho_{\mathcal{E}_{p}}$. If the matrix $\phi=\mathrm{Tr}_{2}\sqrt{J^{\dag}%
J}=\mathrm{Tr}_{2}\sqrt{JJ^{\dag}}$ is scalar (i.e., both of its
eigenvalues are equal), then the trace distance between the Choi
matrices $||J||$ is
equal to the diamond distance between the channels $\Delta_{M}(p)$%
~\cite[Proposition~1]{nechita}. After simple algebra we indeed
find
\begin{equation}
\phi=\frac{\xi_{M}}{8}\left[  2(1-p)+a_{-}+a_{+}\right]
\begin{pmatrix}
1 & 0\\
0 & 1
\end{pmatrix}
,
\end{equation}
where $a_{\pm}=\sqrt{1-p}\sqrt{5\pm4\sqrt{1-p}-p}$. Because $\phi$
is scalar, the condition above is met and the expression of
$\Delta_{M}(p)$ is twice the
(degenerate) eigenvalue of $\phi$, i.e.,%
\begin{equation}
\Delta_{M}(p)=\frac{\xi_{M}}{4}\left[  2(1-p)+a_{-}+a_{+}\right]  ,
\end{equation}
which simplifies to Eq.~(\ref{Deltadamp}).

\subsection*{Aknowledgements}

This work has been supported by the EPSRC via the `UK Quantum
Communications Hub' (EP/M013472/1) and by the European Commission
via `Continuous Variable Quantum Communications' (CiViQ, Project
ID: 820466).~The authors would like to thank Satoshi Ishizaka, Sam
Braunstein, Seth Lloyd, Gaetana Spedalieri, and Zhi-Wei Wang for
feedback.

%

\newpage

\begin{center}
\textbf{{\large Supplementary Information}}
\end{center}

\setcounter{section}{0} 
\setcounter{figure}{0}
\renewcommand{\thefigure}{S\arabic{figure}}
\renewcommand{\bibnumfmt}[1]{[S#1]}
\renewcommand{\citenumfont}[1]{S#1}

\section{Depolarizing probability for qubit PBT}

Here we show the formula for the PBT numbers $\xi_{M}$ given in
Eq.~(11) of the main text. Define the states
\begin{align}
|\phi^{-}\rangle &  =\left(  |01\rangle-|10\rangle\right)  /\sqrt{2},\\
\sigma^{i}  &  =\frac{1}{2^{M-1}}|\phi^{-}\rangle\langle\phi^{-}|_{A_{i}%
C}\otimes\mathbb{I}_{\mathbf{\bar{A}}}^{M-1},\\
\rho &  =\sum_{i=1}^{M}\sigma^{i},
\end{align}
where $\mathbb{I}_{\mathbf{\bar{A}}}^{M-1}$ is the
$2^{M-1}$-dimensional
identity operator acting on the $M-1$ qubits $\mathbf{\bar{A}}=\mathbf{A}%
\backslash A_{i}$ (similarly, we denote
$\mathbf{\bar{B}}=\mathbf{B}\backslash B_{i}$). Then, in
qubit-based PBT with $M$ ports, one uses a POVM with
operators%
\begin{equation}
\Pi_{C\mathbf{A}}^{i}=\rho^{-\frac{1}{2}}\sigma^{i}\rho^{-\frac{1}{2}}%
+M^{-1}\left(  \mathbb{I}^{M+1}-\rho^{-\frac{1}{2}}\rho\rho^{-\frac{1}{2}%
}\right)  ,
\end{equation}
where $\mathbb{I}^{M+1}$ is the $2^{M+1}$-dimensional identity
operator acting on the input qubit $C$ and Alice's resource qubits
$\mathbf{A}$, while $\rho^{-\frac{1}{2}}$ is taken over the
support of $\rho$~\cite{PBT1s}.

Since the resource state%
\begin{equation}
\Phi_{\mathbf{AB}}^{\otimes M}=\bigotimes_{i=1}^{M}|\phi^{-}\rangle_{i}%
\langle\phi^{-}|
\end{equation}
is symmetric under exchange of labels, we can calculate $\xi_{M}$
assuming that the qubit is teleported to the first port, and hence
we only need to consider $\Pi_{C\mathbf{A}}^{1}$. The PBT channel
$\Gamma_{M}$ from qubit $C$ to qubit $B_{1}$ is a depolarizing
channel with isotropic Choi matrix
\begin{equation}
(\rho_{\Gamma_{M}})_{DB_{1}}=%
\begin{pmatrix}
\frac{1}{2}-\frac{\xi_{M}}{4} & 0 & 0 & \frac{1}{2}-\frac{\xi_{M}}{2}\\
0 & \frac{\xi_{M}}{4} & 0 & 0\\
0 & 0 & \frac{\xi_{M}}{4} & 0\\
\frac{1}{2}-\frac{\xi_{M}}{2} & 0 & 0 & \frac{1}{2}-\frac{\xi_{M}}{4}%
\end{pmatrix}
,
\end{equation}
where $\xi_{M}$ is the probability of depolarizing and $D$ is the
ancillary system not passing through the PBT channel. Note that we
can equivalently use
a POVM $\{\Pi_{C\mathbf{A}}^{i}\}$ and a resource state $\Phi_{\mathbf{AB}%
}^{\otimes M}$ where we replace
\begin{equation}
|\phi^{-}\rangle\rightarrow|\Phi\rangle=\left(
|00\rangle+|11\rangle\right) /\sqrt{2}.
\end{equation}

In order to find $\xi_{M}$, it suffices to find any one of the
non-zero elements in the output Choi matrix. Selecting the
coefficient of $|01\rangle_{DB_{1}}\langle01|$, our expression is
\begin{align}
\frac{\xi_{M}}{4M}  &
=\langle01|\mathrm{Tr}_{\mathbf{A\bar{B}}C}\left[
\sqrt{\Pi^{1}}\Phi_{\mathbf{AB}}^{\otimes
M}|\Phi\rangle_{CD}\langle\Phi
|\sqrt{\Pi^{1}}\right]  |01\rangle\\
&  =\mathrm{Tr}\left[
\Pi_{C\mathbf{A}}^{1}\Phi_{\mathbf{AB}}^{\otimes
M}|\Phi\rangle_{CD}\langle\Phi||01\rangle_{DB_{1}}\langle01|\right] \\
&  =\tfrac{1}{2}\mathrm{Tr}\left[  \Pi_{C\mathbf{A}}^{1}\langle1|_{B_{1}%
}\mathrm{Tr}_{\mathbf{\bar{B}}}(\Phi_{\mathbf{AB}}^{\otimes
M})|1\rangle _{B_{1}}\otimes|0\rangle_{C}\langle0|\right]  ,
\end{align}
where the factor of $M$ comes from the fact we have $M$ possible
outcomes and where the third line is due to $|\Phi\rangle_{CD}$.
Considering the structure of $\Phi_{\mathbf{AB}}^{\otimes M}$, we
can write
\begin{equation}
\xi_{M}=\frac{M}{2^{M-1}}\mathrm{Tr}\left[
\Pi_{C\mathbf{A}}^{1}\left(
\mathbb{I}_{\mathbf{\bar{A}}}^{M-1}\otimes|00\rangle_{A_{1}C}\langle
00|\right)  \right]  .
\end{equation}

Ref.~\cite{PBT1s} showed that, by using a spinorial basis, the
eigenvectors and eigenvalues of $\rho$ can be expressed in a
simple form. Defining the basis vectors
$\{|\Phi^{M}(j,m,\alpha)\rangle\}$, where $j$ is the total spin,
$m$ is the spin component in the z-basis and $\alpha$ is a
degeneracy value, they
constructed the eigenvectors of $\rho$ as%
\begin{gather}
|\Psi(\lambda_{j}^{\mp},m,\alpha)\rangle=\\
|\Phi^{M}(j,m+\frac{1}{2},\alpha)\rangle_{\mathbf{A}}|0\rangle_{C}\left\langle
j,m+\frac{1}{2},\frac{1}{2},-\frac{1}{2}\middle|j\pm\frac{1}{2},m\right\rangle
\nonumber\\
+|\Phi^{M}(j,m-\frac{1}{2},\alpha)\rangle_{\mathbf{A}}|1\rangle_{C}%
\left\langle j,m-\frac{1}{2},\frac{1}{2},\frac{1}{2}\middle|j\pm\frac{1}%
{2},m\right\rangle ,\nonumber
\end{gather}
where the terms in the large, triangular brackets are the
Clebsch-Gordan coefficients. They found that these eigenvectors
correspond to the eigenvalues
\begin{equation}
\lambda_{j}^{-}=\frac{1}{2}\left(  \frac{M}{2}-j\right)  ,~\lambda_{j}%
^{+}=\frac{1}{2}\left(  \frac{M}{2}+j+1\right)  . \label{eq:eigenvalues}%
\end{equation}
(Our expressions differ from those in Ref.~\cite{PBT1s} by a
factor of $2^{M+1}$ due to including this factor in the definition
of the $\sigma^{i}$).
Then, Ref.~\cite{PBT1s} expressed the state $\rho$ as%
\begin{align}
\rho &  =\sum_{s,m,\alpha}\lambda_{j}^{-}|\Psi(\lambda_{s-\frac{1}{2}}%
^{-},m,\alpha)\rangle\langle\Psi(\lambda_{s-\frac{1}{2}}^{-},m,\alpha
)|\nonumber\\
&
+\lambda_{j}^{+}|\Psi(\lambda_{s+\frac{1}{2}}^{+},m,\alpha)\rangle
\langle\Psi(\lambda_{s+\frac{1}{2}}^{+},m,\alpha)|. \label{eq:rho}%
\end{align}

Note that the basis vectors on an $M$-spin system,
$\{|\Phi^{M}(j,m,\alpha )\rangle\}$, can be divided into two types
based on how they are constructed
from the basis vectors on an $(M-1)$-spin system, $\{|\Phi^{M-1}%
(j,m,\alpha)\rangle\}$. Specifically%
\begin{gather}
|\Phi_{I}^{M}(j,m)\rangle=\\
|\Phi^{M-1}(j+\frac{1}{2},m+\frac{1}{2})\rangle|0\rangle\left\langle
j+\frac{1}{2},m+\frac{1}{2},\frac{1}{2},-\frac{1}{2}\middle|j,m\right\rangle
\nonumber\\
+|\Phi^{M-1}(j+\frac{1}{2},m-\frac{1}{2})\rangle|1\rangle\left\langle
j+\frac{1}{2},m-\frac{1}{2},\frac{1}{2},\frac{1}{2}\middle|j,m\right\rangle
,\nonumber
\end{gather}
and%
\begin{gather}
|\Phi_{II}^{M}(j,m)\rangle=\\
|\Phi^{M-1}(j-\frac{1}{2},m+\frac{1}{2})\rangle|0\rangle\left\langle
j-\frac{1}{2},m+\frac{1}{2},\frac{1}{2},-\frac{1}{2}\middle|j,m\right\rangle
\nonumber\\
+|\Phi^{M-1}(j-\frac{1}{2},m-\frac{1}{2})\rangle|1\rangle\left\langle
j-\frac{1}{2},m-\frac{1}{2},\frac{1}{2},\frac{1}{2}\middle|j,m\right\rangle
,\nonumber
\end{gather}
where we have omitted the label $\alpha$, but both component
vectors are assumed to have the same degeneracy value. We also
divide the eigenvectors
$\{|\Psi(\lambda_{j}^{\mp},m,\alpha)\rangle\}$ into types $I$ and
$II$, based
on whether they are constructed from the vectors $\{|\Phi_{I}^{M}%
(j,m,\alpha)\rangle\}$ or $\{|\Phi_{II}^{M}(j,m,\alpha)\rangle\}$.
Ref.~\cite{PBT1s} also used the explicit forms of the
Clebsch-Gordon
coefficients to calculate the expressions%
\begin{gather}
\langle\phi^{-}|_{A_{1}C}|\Psi_{I}(\lambda_{s}^{-}-\frac{1}{2},m,\alpha
)\rangle_{\mathbf{A}C}=\nonumber\\
\sqrt{\frac{s}{2s+1}}|\Phi^{M-1}(s,m,\alpha)\rangle_{\mathbf{\bar{A}}%
}\label{expression 1}\\
\langle\phi^{-}|_{A_{1}C}|\Psi_{I}(\lambda_{s}^{+}+\frac{1}{2},m,\alpha
)\rangle_{\mathbf{A}C}=0,\\
\langle\phi^{-}|_{A_{1}C}|\Psi_{II}(\lambda_{s}^{-}-\frac{1}{2},m,\alpha
)\rangle_{\mathbf{A}C}=0,\\
\langle\phi^{-}|_{A_{1}C}|\Psi_{II}(\lambda_{s}^{+}+\frac{1}{2},m,\alpha
)\rangle_{\mathbf{A}C}=\nonumber\\
-\sqrt{\frac{s+1}{2s+1}}|\Phi^{M-1}(s,m,\alpha)\rangle_{\mathbf{\bar{A}}}.
\label{expression 4}%
\end{gather}

We can express $\sigma^{1}$as%
\begin{gather}
\sigma^{1}=|\phi^{-}\rangle_{A_{1}C}\langle\phi^{-}|\otimes\nonumber\\
\sum_{j=j_{min}}^{\frac{M-1}{2}}\sum_{m=-j}^{j}\sum_{\alpha}|\Phi
^{M-1}(j,m,\alpha)\rangle_{\mathbf{\bar{A}}}\langle\Phi^{M-1}(j,m,\alpha)|
\label{eq:sigma}%
\end{gather}
where the term over the $\mathbf{\bar{A}}$ qubits is the identity.
We now separate out the contributions from the two terms of
$\Pi^{1}$, writing
\begin{align}
\Pi^{1}  &  =\pi_{0}+\pi_{1},\\
\pi_{0}  &  =M^{-1}\left(
\mathbb{I}^{M+1}-\rho^{-\frac{1}{2}}\rho
\rho^{-\frac{1}{2}}\right)  ,\\
\pi_{1}  &  =\rho^{-\frac{1}{2}}\sigma^{1}\rho^{-\frac{1}{2}}.
\end{align}
Here $\pi_{0}$ is simply $M^{-1}$ times the identity over the
vector space that does not lie in the support of $\rho$, and
corresponds to those eigenvectors of $\rho$ with eigenvalue 0,
namely $\{|\Psi_{II}(\lambda _{\frac{M}{2}}^{-},m)\rangle\}$
(omitting the label $\alpha$, since the degeneracy is $1$ for this
choice of $j$). Consequently, we may write
\begin{equation}
\pi_{0}=M^{-1}\sum_{m=-\frac{M+1}{2}}^{\frac{M+1}{2}}|\Psi_{II}(\lambda
_{\frac{M}{2}}^{-},m)\rangle\langle\Psi_{II}(\lambda_{\frac{M}{2}}^{-},m)|.
\end{equation}
Combining the expressions for $\rho$ and $\sigma^{1}$ in
Eqs.~(\ref{eq:rho})
and (\ref{eq:sigma}) and the expressions in Eqs.~(\ref{expression 1}%
)-(\ref{expression 4}), we can write%
\begin{align}
\pi_{1}  &  =\sum_{s=s_{min}}^{(M-1)/2}\sum_{m=-s}^{s}\sum_{\alpha}\\
&  \left[  (\lambda_{s-\frac{1}{2}}^{-})^{-1}\frac{s}{2s+1}|\Psi_{I}%
(\lambda_{s-\frac{1}{2}}^{-},m,\alpha)\rangle\langle\Psi_{I}(\lambda
_{s-\frac{1}{2}}^{-},m,\alpha)|\right. \nonumber\\
&  -(\lambda_{s-\frac{1}{2}}^{-}\lambda_{s+\frac{1}{2}}^{+})^{-\frac{1}{2}%
}\frac{\sqrt{s(s+1)}}{2s+1}\nonumber\\
&  \left(
|\Psi_{I}(\lambda_{s-\frac{1}{2}}^{-},m,\alpha)\rangle\langle
\Psi_{II}(\lambda_{s+\frac{1}{2}}^{+},m,\alpha)|+\right. \nonumber\\
&  \left.
|\Psi_{II}(\lambda_{s+\frac{1}{2}}^{+},m,\alpha)\rangle\langle
\Psi_{I}(\lambda_{s-\frac{1}{2}}^{-},m,\alpha)|\right)  +\nonumber\\
&  \left.  (\lambda_{s+\frac{1}{2}}^{+})^{-1}\frac{s+1}{2s+1}|\Psi
_{II}(\lambda_{s+\frac{1}{2}}^{+},m,\alpha)\rangle\langle\Psi_{II}%
(\lambda_{s+\frac{1}{2}}^{+},m,\alpha)|\right]  ,\nonumber
\end{align}
where $s_{min}=1/2$ for even $M$, and $0$ for odd $M$.

By calculating the Clebsch-Gordan coefficients, we find%
\begin{gather}
\langle\Psi_{I}(\lambda_{s-\frac{1}{2}}^{-},m)|(\mathbb{I}^{M-1}%
\otimes|00\rangle\langle00|)|\Psi_{I}(\lambda_{s-\frac{1}{2}}^{-}%
,m)\rangle=\nonumber\\
\frac{(s-m)(s+m+1)}{2s(2s+1)},\label{contraction1}\\
\langle\Psi_{II}(\lambda_{s+\frac{1}{2}}^{+},m)|(\mathbb{I}^{M-1}%
\otimes|00\rangle\langle00|)|\Psi_{II}(\lambda_{s+\frac{1}{2}}^{+}%
,m)\rangle=\nonumber\\
\frac{(s-m)(s+m+1)}{2(s+1)(2s+1)},\\
\langle\Psi_{I}(\lambda_{s-\frac{1}{2}}^{-},m)|(\mathbb{I}^{M-1}%
\otimes|00\rangle\langle00|)|\Psi_{II}(\lambda_{s+\frac{1}{2}}^{+}%
,m)\rangle=\nonumber\\
\frac{(s-m)(s+m+1)}{2(2s+1)\sqrt{s(s+1)}},\label{contraction3}\\
\langle\Psi_{II}(\lambda_{\frac{M}{2}}^{-},m)|(\mathbb{I}^{M-1}\otimes
|00\rangle\langle00|)|\Psi_{II}(\lambda_{\frac{M}{2}}^{-},m)\rangle
=\nonumber\\
\frac{M-1-2m}{2M}\left(  \frac{1}{2}-\frac{m}{M+1}\right)  . \label{cont4}%
\end{gather}
Using Eq.~(\ref{cont4}) and summing over $m$, we find
\begin{equation}
\frac{M}{2^{M-1}}\mathrm{Tr}\left[  \pi_{0}\left(
\mathbb{I}_{\mathbf{\bar
{A}}}^{M-1}\otimes|01\rangle_{A_{1}C}\langle01|\right)  \right]  =\frac{1}%
{3}\frac{M+2}{2^{M-1}}. \label{vv1}%
\end{equation}
Using Eqs.~(\ref{contraction1})-(\ref{contraction3}), we find%
\begin{gather}
\frac{M}{2^{M-1}}\mathrm{Tr}\left[  \pi_{1}\left(
\mathbb{I}_{\mathbf{\bar
{A}}}^{M-1}\otimes|01\rangle_{A_{1}C}\langle01|\right)  \right]  =\nonumber\\
\frac{M}{2^{M-1}}\sum_{s=s_{min}}^{(M-1)/2}\sum_{m=-s}^{s}\sum_{\alpha}%
\frac{(s-m)(s+m+1)}{2(2s+1)^{2}}\nonumber\\
\times\left[  (\lambda_{s-\frac{1}{2}}^{-})^{-1}-2(\lambda_{s-\frac{1}{2}}%
^{-}\lambda_{s+\frac{1}{2}}^{+})^{-\frac{1}{2}}+(\lambda_{s+\frac{1}{2}}%
^{+})^{-1}\right]  . \label{eq40}%
\end{gather}

We can simplify the term on the RHS, using%
\begin{align}
&  \left[  (\lambda_{s-\frac{1}{2}}^{-})^{-1}-2(\lambda_{s-\frac{1}{2}}%
^{-}\lambda_{s+\frac{1}{2}}^{+})^{-\frac{1}{2}}+(\lambda_{s+\frac{1}{2}}%
^{+})^{-1}\right] \nonumber\\
&  =\left[
(\lambda_{s-\frac{1}{2}}^{-})^{-\frac{1}{2}}-(\lambda_{s+\frac
{1}{2}}^{+})^{-\frac{1}{2}}\right]  ^{2}.
\end{align}
The degeneracy \textrm{g}$[s]$ for a given $s$-value is given by
\begin{equation}
\mathrm{g}[s]=\frac{(2s+1)(M-1)!}{\left(  \frac{M-1}{2}-s\right)
!\left( \frac{M+1}{2}+s\right)  !},
\end{equation}
and substituting this into Eq.~(\ref{eq40}), we get%
\begin{gather}
\frac{M}{2^{M-1}}\mathrm{Tr}\left[  \pi_{1}\left(
\mathbb{I}_{\mathbf{\bar
{A}}}^{M-1}\otimes|01\rangle_{A_{1}C}\langle01|\right)  \right]  =\nonumber\\
\sum_{s=s_{min}}^{(M-1)/2}\sum_{m=-s}^{s}\frac{(s-m)(s+m+1)}{2^{M}%
(2s+1)}\binom{M}{\frac{M-1}{2}-s}\nonumber\\
\times8\frac{\left(  M+2\right)  -\sqrt{\left(  M+2\right)
^{2}-\left( 2s+1\right)  ^{2}}}{\left(  M+2\right)  ^{2}-\left(
2s+1\right)  ^{2}},
\end{gather}
where we have substituted in the expressions from
Eq.~(\ref{eq:eigenvalues}).
Summing over $m$, we get%
\begin{gather}
\frac{M}{2^{M-1}}\mathrm{Tr}\left[  \pi_{1}\left(
\mathbb{I}_{\mathbf{\bar
{A}}}^{M-1}\otimes|01\rangle_{A_{1}C}\langle01|\right)  \right]  =\nonumber\\
\sum_{s=s_{min}}^{(M-1)/2}\frac{1}{3}\frac{s(s+1)}{2^{M-4}}\binom{M}%
{\frac{M-1}{2}-s}\nonumber\\
\times\frac{\left(  M+2\right)  -\sqrt{\left(  M+2\right)
^{2}-\left( 2s+1\right)  ^{2}}}{\left(  M+2\right)  ^{2}-\left(
2s+1\right)  ^{2}}.
\label{vv2}%
\end{gather}
Therefore, by combining Eqs.~(\ref{vv1}) and (\ref{vv2}), we
finally get the expression of $\xi_{M}$ given in Eq.~(11) of the
main text. We can numerically verify that $\xi_{M}$ scales as
$M^{-1}$ for large $M$.

One may check that Eq.~(11) of the main text can equivalently be
obtained by combining Eq.~(47) of our Methods section (i.e.,
Eq.~(8) of our Lemma~1) together with the expression of the
entanglement fidelity for qubit-based PBT which is given in Eq.
(29) of Ref.~\cite{PBT1s}. For completeness we report this
algebraic check here.

We start from the expression of $\xi_{M}$ given in Eq.~(11) of the
main text. By substituting this expression in Eq.~(10) of the main
text for $d=2$, we get
\begin{align}
\begin{split}
\delta_M&=\frac{3}{2}\xi_{M}\\
&=\frac{M+2}{2^{M}}+\sum_{s=s_{min}%
}^{(M-1)/2}\frac{s(s+1)}{2^{M-3}}\binom{M}{\frac{M-1}{2}-s}\times\\
& \frac{\left(  M+2\right)  -\sqrt{\left(  M+2\right)  ^{2}-\left(
2s+1\right)  ^{2}}}{\left(  M+2\right)  ^{2}-\left(  2s+1\right)
^{2}}.
\end{split}
\end{align}
We now use the fact that each term in the sum would be the same if
we set $s$ to $-(s+1)$, and write
\begin{align}
\begin{split}
\delta_M&=\frac{M+2}{2^{M}}+\sum_{s=-(M+1)/2%
}^{(M-1)/2}\frac{s(s+1)}{2^{M-2}}\binom{M}{\frac{M-1}{2}-s}\times\\
& \frac{\left(  M+2\right)  -\sqrt{\left(  M+2\right)  ^{2}-\left(
2s+1\right)  ^{2}}}{\left(  M+2\right)  ^{2}-\left(  2s+1\right)
^{2}}.
\end{split}
\end{align}
Then, we carry out a change of variables, substituting in
$k=\frac{M-1}{2}-s$, and get
\begin{align}
\begin{split}
\delta_M&=\frac{M+2}{2^{M}}+\sum_{k=0%
}^{M}\frac{(M-2k-1)(M-2k+1)}{2^{M}}\binom{M}{k}\times\\
& \frac{\left(  M+2\right)  -\sqrt{\left(  M+2\right)  ^{2}-\left(
M-2k\right)  ^{2}}}{\left(  M+2\right)  ^{2}-\left(  M-2k\right)
^{2}}.
\end{split}
\end{align}
Using
\begin{align}
(M+2)^{2}-(M-2k)^{2}=4(M-k+1)(k+1),
\end{align}
we write
\begin{align}
\begin{split}
\delta_M&=\frac{M+2}{2^{M}}+\sum_{k=0%
}^{M}\frac{(M-2k)^2-1}{2^{M+2}}\binom{M}{k}\times\\
&\frac{(M+2)-2\sqrt{(M-k+1)(k+1)}}{(M-k+1)(k+1)}.
\end{split}\label{deltapro}
\end{align}

We now use the expression for the entanglement fidelity of qubit
PBT given in~\cite{PBT1s}, i.e.,
\begin{align}
f_{e}=\frac{1}{2^{M+3}}\sum_{k=0}^M
\left(\frac{M-2k-1}{\sqrt{k+1}}+\frac{M-2k+1}{\sqrt{M-k+1}}
\right)^2 \binom{M}{k}.\label{PBTfe}
\end{align}
Combining this with Eq.~(\ref{deltapro}) and expanding the term in
brackets, we can write
\begin{align}
\begin{split}
&f_{e}+\frac{\delta_M}{2}=\frac{M+2}{2^{M+1}}+\sum_{k=0}^{M}\frac{2^{-(M+3)}}{(M-k+1)(k+1)}\binom{M}{k}\times\\
&\left[\left((M-2k)^2-1\right)\left((M+2)-2\sqrt{(M-k+1)(k+1)}\right)\right.\\
&\left.+\left(M+2+M(M-2k)^2+2\left((M-2k)^2-1\right)\times\right.\right.\\
&\left.\left.\sqrt{(M-k+1)(k+1)}\right)\right].
\end{split}
\end{align}
Algebraically simplifying the term in the square brackets, we get
\begin{equation}
f_{e}+\frac{\delta_M}{2}=\frac{M+2}{2^{M+1}}+\sum_{k=0}^{M}\frac{(M-2k)^2}{2^{M+2}(M+2)}\binom{M+2}{k+1},
\end{equation}
and changing variables again, substituting in $x=k+1$, we can
write
\begin{align}
f_{e}+\frac{\delta_M}{2}&=\sum_{x=0}^{M+2}\frac{(M+2-2x)^2}{2^{M+2}(M+2)}\binom{M+2}{x},\label{eq:F+deltaM}
\end{align}
where we have split the term outside the sum into the
contributions for the $x=0$ and $x=M+2$ cases. We now use the
known sums of binomial coefficients,
\begin{align}
\begin{split}
&\sum_{x=0}^{n}\binom{n}{x}=2^n,~~~\sum_{x=0}^{n}x\binom{n}{x}=n2^{n-1},\\
&\sum_{x=0}^{n}x^2\binom{n}{x}=(n+n^2)2^{n-2},
\end{split}
\end{align}
and split the sum in Eq.~(\ref{eq:F+deltaM}) into contributions
from $(M+2)^2$, $(2x)(M+2)$ and $(2x)^2$, getting
\begin{align}
\begin{split}
f_{e}+\frac{\delta_M}{2}&=\frac{2^{-(M+2)}}{M+2}\left(2^{M+2}(M+2)^2-2^{M+3}(M+2)^2\right.\\
&\left.+2^{M+2}\left((M+2)+(M+2)^2\right)\right),
\end{split}
\end{align}
which cancels to give $f_{e}+\frac{\delta_M}{2}=1$, in agreement
with $\delta_M=2(1-f_{e})$. This check is equivalent to say that,
by using Eq.~(\ref{PBTfe}) together with Eq.~(8) of the main text,
we can equivalently obtain Eq.~(11) of the main text (specifically
for qubits).


\section{Ultimate single-photon quantum optical
resolution\label{AppRESOLUTION}}

Consider the problem of discriminating between the following
situations:

\begin{description}
\item[(1)] A point-like source emitting light from position $x=s/2$;

\item[(2)] A point-like source emitting light from the shifted position
$x=-s/2$.
\end{description}

\noindent The discrimination is achieved by measuring the image
created by a focusing optical system. More precisely, we consider
a linear imaging system in the paraxial approximation that is used
to image point-like sources. This is characterized by the Fresnel
number
\begin{equation}
\mathcal{F}=\frac{\ell}{x_{R}}\,,
\end{equation}
where $\ell$ is the size of the object, and
\begin{equation}
x_{R}=\frac{\lambda}{N_{A}}%
\end{equation}
is the Rayleigh length. Here $\lambda$ is the wavelength and
$N_{A}=R/D$ is the numerical aperture, where $R$ is the radius of
the pupil and $D$ is the distance from the object. In the
far-field regime, light is attenuated by a loss parameter
$\eta\simeq\mathcal{F}$ \cite{Shap,m1,m2}. In particular, because
we consider point-like sources, we are in the regime $\eta\ll1$.

First we need to model the imaging system as a quantum channel
acting on the \textit{input} state represented by the light
emitted by the source. The two cases are described by the
following Heisenberg-picture transformations on the input
annihilation operator $a$
\begin{align}
(1)  &  :~a\rightarrow\sqrt{\eta}b_{1}+\sqrt{1-\eta}v_{1},\\
(2)  &  :~a\rightarrow\sqrt{\eta}b_{2}+\sqrt{1-\eta}v_{2},
\end{align}
where $b_{1,2}$ are the output operators (encoding the position of
the source) and $v_{1,2}$ are associated with a vacuum
environment. The modes $b_{1}$, $b_{2}$ are defined on the image
plane and have the form
\begin{equation}
b_{j}=\int dx\,\psi_{j}(x)\,a(x)\,,
\end{equation}
where $a(x),a(x)^{\dag}$ is a continuous family of canonical
operators $[a(x),a(y)^{\dag}]=\delta(x-y)$ defined on the image
plane (for simplicity, we assume unit magnification factor). In
general the image modes $b_{1}$, $b_{2}$ do satisfy the
(non-canonical) commutation relations
\begin{equation}
\lbrack b_{1},b_{2}^{\dag}]=\int
dx\,\psi_{1}(x)\psi_{2}^{\ast}(x)\,,
\end{equation}
where $\psi_{j}$ is the point-spread function associated to the
source being
in poisiton $j$. Then, by setting $\delta=\mathrm{Re}\int dx\,\psi_{1}%
(x)\psi_{2}^{\ast}(x)$, we can define the effective image operators%
\begin{equation}
b_{\pm}:=(b_{1}\pm b_{2})/\sqrt{2(1\pm\delta)}.
\end{equation}
The fact that $\delta\neq0$ means that the two image fields
overlap and the sources cannot be perfectly distinguished. This is
a manifestation of diffraction through the finite objective of the
optical imaging system.

As a result, we can write the action of the channels as
\begin{align}
(1)  &
:~a\rightarrow\sqrt{\eta_{+}}b_{+}+\sqrt{\eta_{-}}b_{-}+\sqrt{1-\eta
}v_{1},\label{ch1}\\
(2)  &
:~a\rightarrow\sqrt{\eta_{+}}b_{+}-\sqrt{\eta_{-}}b_{-}+\sqrt{1-\eta
}v_{2}, \label{ch2}%
\end{align}
where $\eta_{\pm}:=(1\pm\delta)\eta/2$. For simplicity, consider a
single-photon state at the input. We then have
\begin{align}
(1):~|1\rangle &  \rightarrow\eta|\psi_{+}\rangle\langle\psi_{+}%
|+(1-\eta)|0\rangle\langle0|\,,\\
(2):~|1\rangle &  \rightarrow\eta|\psi_{-}\rangle\langle\psi_{-}%
|+(1-\eta)|0\rangle\langle0|\,,
\end{align}
where
\begin{equation}
|\psi_{\pm}\rangle=\frac{\sqrt{\eta_{+}}|1\rangle_{+}-\sqrt{\eta_{-}}%
|1\rangle_{-}}{\sqrt{\eta}}~.
\end{equation}
More generally, the action of the channels on a generic pure input
state is given by
\begin{align}
(1):~\alpha|0\rangle+\beta|1\rangle &  \rightarrow\left(  |\alpha|^{2}%
+\eta|\beta|^{2}\right)  \times\label{teleg}\\
&
|\psi_{+}(\alpha,\beta)\rangle\langle\psi_{+}(\alpha,\beta)|+(1-\eta
)|\beta|^{2}|0\rangle\langle0|,\nonumber\\
(2):~\alpha|0\rangle+\beta|1\rangle &  \rightarrow\left(  |\alpha|^{2}%
+\eta|\beta|^{2}\right)  \times\label{teleg2}\\
&
|\psi_{-}(\alpha,\beta)\rangle\langle\psi_{-}(\alpha,\beta)|+(1-\eta
)|\beta|^{2}|0\rangle\langle0|,\nonumber
\end{align}
where
\begin{equation}
|\psi_{\pm}(\alpha,\beta)\rangle=\frac{\alpha|0\rangle+\beta\sqrt{\eta_{+}%
}|1\rangle_{+}\pm\beta\sqrt{\eta_{-}}|1\rangle_{-}}{\sqrt{|\alpha|^{2}%
+\eta|\beta|^{2}}}~.
\end{equation}
As we can see from Eqs.~(\ref{teleg}) and~(\ref{teleg2}), if we
apply a Pauli operator $X$~\cite{NiChs} to the input state
$\alpha|0\rangle+\beta|1\rangle$, we have a swap between $\alpha$
and $\beta$. This leads to an output state with a different
eigenspectrum, so that it cannot be obtained by applying a
unitary. This means that the quantum channels are \textit{not}
teleportation-covariant.

By limiting ourselves to the space of either no photon or one
photon $\mathcal{H}_{2}=\mathrm{span}\{|0\rangle,|1\rangle\}$, the
the input space of the channels is a qubit, and their output is a
qutrit, so that the dimension of the input Hilbert space is $d=2$.
Apart from restricting the input space to qubits, we assume the
most general adaptive strategy allowed by quantum mechanics, so
that the quantum state of the source may be optimized as a
consequence of the output (as generally happens in the adaptive
protocol discussed in the main text). In order to compute the
ultimate performance, we need to compute the quantum fidelity
between the Choi matrices of the two channels in Eqs.~(\ref{ch1})
and~(\ref{ch2}) suitably truncated to $\mathcal{H}_{2}$.

Consider then the maximally entangled state
$|\Phi_{2}\rangle=(|0\rangle
|1\rangle+|1\rangle|0\rangle)/\sqrt{2}$. The Choi matrices
associated with the two truncated channels are equal to
\begin{align}
\rho_{(1)}  &
=\frac{1+\eta}{2}|\Psi_{+}\rangle\langle\Psi_{+}|+\frac{1-\eta
}{2}|0\rangle\langle0|\,,\\
\rho_{(2)}  &
=\frac{1+\eta}{2}|\Psi_{-}\rangle\langle\Psi_{-}|+\frac{1-\eta
}{2}|0\rangle\langle0|\,,
\end{align}
where
\begin{equation}
|\Psi_{\pm}\rangle=\frac{|0\rangle|1\rangle+\sqrt{\eta_{+}}|1\rangle
_{+}|0\rangle\pm\sqrt{\eta_{-}}|1\rangle_{-}|0\rangle}{\sqrt{1+\eta}}.
\end{equation}
Notice that
$\langle\Psi_{+}|\Psi_{-}\rangle=(1+\delta\eta)/(1+\eta)$ where
$\delta\eta:=\eta_{+}-\eta_{-}$. Therefore we obtain the fidelity
\begin{align}
F(\rho_{(1)},\rho_{(2)})  &
=\mathrm{Tr}\sqrt{\sqrt{\rho_{(1)}}\,\rho
_{(2)}\,\sqrt{\rho_{(1)}}}\\
&  =\frac{1+\eta}{2}\left\vert
\frac{1+\delta\eta}{1+\eta}\right\vert
+\frac{1-\eta}{2}=\frac{1-\eta+|1+\delta\eta|}{2}.\nonumber
\end{align}
Assuming that $\delta$ is real, this becomes%
\begin{equation}
F[\rho_{(1)},\rho_{(2)}]=1-\frac{\eta(1-\delta)}{2},
\end{equation}
which allows us to identify $\epsilon=\eta(1-\delta)/2$. A common
way to model diffraction is to consider a Gaussian point-spread
function, i.e.
\begin{equation}
\psi_{j}(s)\simeq e^{-(x-x_{j})^{2}/4},
\end{equation}
where $x_{j}$ is the center of the $j$th emitter, and the variance
of the Gaussian is $1$ in units of Rayleigh length. Under this
Gaussian model one obtains~\cite{Tsang15s,Lupo16s}
\begin{equation}
\delta\simeq e^{-s^{2}/8},
\end{equation}
where $s$ is the separation in unit of wavelength. Therefore
\begin{equation}
\epsilon\simeq\frac{\eta(1-e^{-s^{2}/8})}{2}\simeq\frac{\eta
s^{2}}{16}.
\end{equation}
By replacing this quantity in Eq.~(27) of the main text with $d=2$
we obtain
the lower bound%
\begin{equation}
B\gtrsim\frac{1}{4}\exp{\left(  -2ns\sqrt{\eta}\right)  }~.
\end{equation}

\section{Ultimate limit of adaptive quantum illumination\label{QIapp}}

\subsection{Standard (non-adaptive) protocol}

In quantum illumination~\cite{Qill0s,Qill1s,Qill2s,Qill3s}, we aim
at determining whether a low-reflectivity object is present or not
in a region with thermal noise. We therefore prepare a signal
system $s$ and an idler system $i$ in a joint entangled state
$\rho_{si}$. The signal system is sent to probe the target while
the idler system is retained for its measurement together with the
potential signal reflection from the target. If the object is
absent, the \textquotedblleft reflected\textquotedblright\ system
is just thermal background noise. If the object is present, then
this is composed of the actual reflection of the signal from the
target plus thermal background noise. This object can be modelled
by a beam splitter, with very small transmissivity $\eta\ll1$,
which combines the each incoming optical mode (signal system) with
a thermal mode with $b$ mean number of photons.

In the discrete-variable version of quantum
illumination~\cite{Qill0s}, the signal system is prepared in an
ensemble of $d$ optical modes, with $1$ photon in one of the modes
and vacuum in the others. This is the number of modes which are
distinguished by the detector in each detection process. If we
introduce the following $d-$dimensional computational basis%
\begin{align}
\left\vert 1\right\rangle  &  :=\overset{d}{\overbrace{\left\vert
00\ldots01\right\rangle }},\\
\left\vert 2\right\rangle  &  :=\left\vert 00\ldots10\right\rangle ,\\
&  \vdots\nonumber\\
\left\vert d-1\right\rangle  &  :=\left\vert 01\ldots00\right\rangle ,\\
\left\vert d\right\rangle  &  :=\left\vert 10\ldots00\right\rangle
,
\end{align}
then the entangled signal-idler state can be written as%
\begin{equation}
\psi_{si}=|\psi\rangle_{si}\left\langle \psi\right\vert
,~~|\psi\rangle
_{si}=d^{-1/2}\sum_{k=1}^{d}|kk\rangle_{si}~. \label{psi}%
\end{equation}

Let us define the $d$-dimensional identity operator $\mathbb{I}^{d}%
:=\sum_{k=1}^{d}|k\rangle\left\langle k\right\vert $ which
projects onto the subspace spanned by the $1$-photon states, and
the $(d+1)$-dimensional identity operator
$\mathbb{I}^{d+1}:=\sum_{k=0}^{d}|k\rangle\left\langle
k\right\vert $ which also includes the vacuum state $\left\vert
0\right\rangle
:=\left\vert 00\ldots00\right\rangle $. Then, we have the reduced idler state%
\begin{equation}
\psi_{i}:=\mathrm{Tr}_{s}(\psi_{si})=d^{-1}\mathbb{I}_{i}^{d},
\end{equation}
and we define the thermal state of the environment as~\cite{Qill0s}%
\begin{equation}
\rho^{\text{th}}(b):=(1-db)|0\rangle\langle0|+b\mathbb{I}^{d},
\end{equation}
where $b$ is the mean number of thermal photons per mode. Here
$b\ll1$ and $db\ll1$, where $db$ is the mean number of thermal
photons in each detection event.

The output $(d+1)\times d$ state of the reflected signal and
retained idler is
given by%
\begin{equation}%
\begin{array}
[c]{rl}%
\text{Target absent:} & ~\sigma=\rho^{\text{th}}(b)\otimes d^{-1}%
\mathbb{I}_{i}^{d},\\
\text{Target present:} & ~\rho=(1-\eta)\sigma+\eta\psi_{si}.
\end{array}
\end{equation}
If the target is probed $n$ times, then we may use the QCB to
bound $Q$ the error probability $p_{\text{err}}$ in the
discrimination of $\rho$ and $\sigma$. In the regime of
signal-to-noise-ratio $\eta d/b\lesssim1$, one finds~\cite{Qill0s}
\begin{equation}
Q=1-\frac{\eta^{2}d}{8b}+\mathcal{O}(b^{2},\eta b)~, \label{QCBappp}%
\end{equation}
which tightens the QCB by a factor $d$ with respect to the
unentangled case where $Q\approx1-\eta^{2}/(8b)$. From
Eq.~(\ref{QCBappp}), we may write the following bound for the
error probability of target detection after $n$
probings~\cite{Qill0s}%
\begin{equation}
p_{n}(\sigma\neq\rho)\leq\frac{1}{2}\exp\left(
-\frac{\eta^{2}dn}{8b}\right) .
\end{equation}
In particular, for $\eta d/b\simeq1$, this can be written as%
\begin{equation}
p_{n}(\sigma\neq\rho)\leq\frac{1}{2}\exp\left(  -\frac{\eta
n}{8}\right)  .
\end{equation}
Note that, for the unentangled case, in the same regime
$\eta/b\simeq1/d$ we may write
$p_{n}\leq\frac{1}{2}\exp[-n\eta/(8d)]$.

\subsection{Adaptive protocol}

The adaptive formulation of the discrete variable protocol of
quantum illumination assumes an unlimited quantum computer with
two register $\mathbf{a}$ and $\mathbf{b}$, prepared in an
arbitrary joint quantum state. In each probing, a system $a$ is
picked from the input register $\mathbf{a}$ and sent to the
target. Its reflection $a^{\prime}$ is stored in the output
register $\mathbf{b}$. A adaptive quantum operation (QO) is
applied to both the update registers before the next transmission
and so on. Therefore any probing is interleaved by the application
of adaptive QOs $\Lambda$'s to the registers, defining the
adaptive protocol $\mathcal{P}_{n}$ (see also the main text for
this description). After $n$ probings, the state of the registers
is $\rho_{n}(u)$ where $u=0,1$ is a bit encoding the absence or
presence of the target. This state is optimally measured by an
Helstrom POVM. By optimizing over all protocol $\mathcal{P}_{n}$,
we define the minimum error probability $p_{n}$ for adaptive
quantum illumination.

Following the constraints and typical regime of DV quantum
illumination, we assume that the signal systems are
$(d+1)$-dimensional qudits described by a basis $\{\left\vert
0\right\rangle ,\left\vert 1\right\rangle ,\ldots ,\left\vert
d\right\rangle \}$, where $\left\vert i\right\rangle :=\left\vert
0\cdots010\cdots0\right\rangle $ has one photon in the $i$th mode.
For this reason, the two possible quantum illumination channels,
$\mathcal{E}_{0}$ and $\mathcal{E}_{1}$, are $(d+1)$-dimensional
channels. In particular, consider as their input the
maximally-entangled state
\begin{equation}
\Psi_{si}=\frac{1}{d+1}\sum_{k,j=0}^{d}|kk\rangle_{si}\langle jj|,
\end{equation}
which is similar to $\psi_{si}$ in Eq.~(\ref{psi}) but also
includes the vacuum state. Then, we may write the following two
$(d+1)\times(d+1)$
dimensional Choi matrices%
\begin{equation}%
\begin{array}
[c]{rl}%
\text{Target absent:} & ~\sigma:=\rho_{\mathcal{E}_{0}}=\rho^{\text{th}%
}(b)\otimes(d+1)^{-1}\mathbb{I}_{i}^{d+1},\\
\text{Target present:} &
~\rho:=\rho_{\mathcal{E}_{1}}=(1-\eta)\sigma+\eta \Psi_{si}.
\end{array}
\end{equation}
It is clear that $\mathcal{E}_{0}$ and $\mathcal{E}_{1}$ are not
jointly teleportation-covariant due to the fact that they have
different transmissivities ($\eta_{0}=0$ and $\eta_{1}=\eta$).

To bound $p_{n}$ we apply Theorem~3 of the main text and, more
specifically, Eq.~(27) of the main text, because $\eta\ll1$ and,
therefore, the fidelity between the Choi matrices can be expanded
as $F(\sigma,\rho)\simeq1-\epsilon$. Thus, let us start by
computing this fidelity. Let us set $x=\sqrt{1-bd}$ and note that
we may write
\begin{equation}
\sqrt{\sigma}=(x|0\rangle_{s}\langle0|+\sqrt{b}\mathbb{I}_{s}^{d}%
)\otimes(d+1)^{-1/2}\mathbb{I}_{i}^{d+1}.
\end{equation}
Then, we may compute
\begin{align}
\Omega^{2}  &  :=\sqrt{\sigma}\rho\sqrt{\sigma}\nonumber\\
&  =\frac{1}{(d+1)^{2}}\left\{  (1-\eta)\left[  x^{4}|0\rangle_{s}%
\langle0|+b^{2}\mathbb{I}_{s}^{d}\right]
\otimes\mathbb{I}_{i}^{d+1}\right.
\nonumber\\
&  +\eta\left[
x^{2}|00\rangle_{si}\langle00|+\sqrt{b}x\sum_{k=1}^{d}\left(
|00\rangle_{si}\langle kk|+|kk\rangle_{si}\langle00|\right)
\right.
\nonumber\\
&  \left.  \left.  +b\sum_{j,k=1}^{d}|kk\rangle_{si}\langle
jj|\right] \right\}  .
\end{align}
One can check that $\Omega^{2}$ has $d^{2}$ degenerate eigenvalues
equal to
$b^{2}(d+1)^{-2}$, $d$ degenerate eigenvalues equal to $(1-\eta)x^{4}%
(d+1)^{-2}$, and other $d+1$ eigenvalues $\{\lambda_{i}\}$ given
by the
diagonalization of the matrix $(d+1)^{-2}\mathbf{M}$ where%
\begin{equation}
\mathbf{M}=\left(
\begin{array}
[c]{ccccc}%
(1-\eta)x^{4}+\eta x^{2} & \eta x\sqrt{b} & \eta x\sqrt{b} &
\cdots & \eta
x\sqrt{b}\\
\eta x\sqrt{b} & b(b+\eta) & \eta b & \cdots & \eta b\\
\eta x\sqrt{b} & \eta b & b(b+\eta) & \ddots & \vdots\\
\vdots & \vdots & \ddots & \ddots & \eta b\\
\eta x\sqrt{b} & \eta b & \cdots & \eta b & b(b+\eta)
\end{array}
\right)  .
\end{equation}

Once we find the eigenvalues of $\Omega^{2}$ we take their square
root so as
to compute those of $\Omega$. Finally, their sum provides $\mathrm{Tr}%
\Omega=F(\sigma,\rho)$. We are interested in the regime of low
thermal noise $b\ll1$ and low reflectivity $\eta\ll1$. There, we
may expand at the leading
orders in $\eta$ and $b$ to get%
\begin{equation}
F(\sigma,\rho)=1-\frac{\eta d+2b-2\sqrt{\eta
db}}{2(d+1)}+\mathcal{O}(\eta
^{2},\eta^{3/2}b^{1/2},\eta b,b^{3/2}). \label{eq1bb}%
\end{equation}
In the typical signal-to-noise-ratio $\eta d/b\simeq1$ of quantum
illumination~\cite{Qill0s}, we may directly re-write
Eq.~(\ref{eq1bb}) as
$F(\sigma,\rho)\simeq1-\epsilon$, where%
\begin{equation}
\epsilon:=\frac{\eta d+2b-2\sqrt{d\eta b}}{2(d+1)}\simeq\frac{d\eta}%
{2(d+1)}<\eta/2,
\end{equation}
up to orders $\mathcal{O}(\eta^{2},\sqrt{\eta b},b)$. By replacing
the latter in Eq.~(27) of the main text (and assuming the correct
dimension $d\rightarrow d+1$), we get the following lower bound
for the minimum error probability
$p_{n}$ of adaptive quantum illumination%
\begin{equation}
p_{n}\geq\frac{1}{4}\exp(-4nd\sqrt{\eta}).
\end{equation}


\section{Adaptive quantum channel estimation}

\subsection{Adaptive protocols for parameter estimation}

As also described in the main text, consider an adaptive protocol
of quantum channel estimation.\ We want to estimate a continuous
parameter $\theta$ encoded in a quantum channel
$\mathcal{E}_{\theta}$ by means of the most general protocols
allowed by quantum mechanics, i.e., based on adaptive QOs as
described in the main text. After $n$ probings, there is a
$\theta$-dependent output state $\rho_{n}(\theta)$ which is
generated by the sequence of QOs
$\{\Lambda_{0},\Lambda_{1},\ldots,\Lambda_{n}\}$ characterizing
the adaptive protocol $\mathcal{P}_{n}$. Finally, the output state
is measured by a POVM $\mathcal{M}$ providing an optimal unbiased
estimator $\tilde{\theta}$ of
parameter $\theta$. The minimum error variance Var$(\tilde{\theta}%
):=\langle(\tilde{\theta}-\theta)^{2}\rangle$ must satisfy the
quantum Cramer-Rao bound (QCRB)~\cite{Sam1s}
\textrm{Var}$(\tilde{\theta})\geq
1/$\textrm{QFI}$_{\theta}(\mathcal{P}_{n})$, where
\textrm{QFI}$_{\theta }(\mathcal{P}_{n})$ is the quantum Fisher
information (QFI) associated with $n$ adaptive uses.

Note that the QFI can be computed as%
\begin{equation}
\mathrm{QFI}_{\theta}(\mathcal{P}_{n})=\frac{4d_{B}^{2}[\rho_{n}(\theta
),\rho_{n}(\theta+d\theta)]}{d\theta^{2}},
\end{equation}
where $d_{B}(\rho,\sigma):=\sqrt{2[1-F(\rho,\sigma)]}$ is the
Bures distance, with $F(\rho,\sigma)$ being the Bures fidelity of
$\rho$ and $\sigma$. The ultimate precision of adaptive quantum
metrology is given by optimizing the QFI over all adaptive
protocols, i.e.,
\begin{equation}
\overline{\text{\textrm{QFI}}}_{\theta}^{n}:=\sup_{\mathcal{P}}\mathrm{QFI}%
_{\theta}(\mathcal{P}_{n}). \label{fisheropt}%
\end{equation}
Contrarily to the cases of sequential or parallel strategies, the
ultimate performance of adaptive quantum metrology is poorly
studied, with limited results for DV programmable channels, and
mainly stated for DV and CV teleportation-covariant channels, such
as Pauli or Gaussian channels~\cite{PirCos}.

\subsection{PBT\ stretching of adaptive quantum metrology}

As shown in Ref.~\cite{PirCos}, the adaptive estimation of a noise
parameter $\theta$ encoded in a teleportation-covariant channel
(i.e., such that the parametrized class of channels
$\mathcal{E}_{\theta}$ is jointly-teleportation covariant) is
limited to the standard quantum limit (SQL). More generally, as
discussed in Ref.~\cite{ReviewMETROs}, the adaptive estimation of
a parameter in a quantum channel cannot beat the SQL if the
channel has a single-copy simulation, i.e., of the type
\begin{equation}
\mathcal{E}_{\theta}(\rho)=\mathcal{S}(\rho\otimes\pi_{\theta}),
\label{simu1copy}%
\end{equation}
where $\mathcal{S}$\ is a (parameter-independent) trace-preserving
QO and $\pi_{\theta}$ is a program state (depending on the
parameter). To beat the SQL, the channel should not admit a
simulation as in Eq.~(\ref{simu1copy}) but
a multi-copy version%
\begin{equation}
\mathcal{E}_{\theta}(\rho)=\mathcal{S}(\rho\otimes\pi_{\theta}^{\otimes
M}),
\end{equation}
for some $M>1$. This is approximately the type of simulation that
we can achieve by using PBT.

First of all, we may replace the channel $\mathcal{E}_{\theta}$
with its
$M$-port approximation $\mathcal{E}_{\theta}^{M}:=\mathcal{E}_{\theta}%
\circ\Gamma_{M}$, where $\Gamma_{M}$ is the $M$-port PBT\ channel.
Using Lemma~1 of the main text, the simulation error may be
bounded as
\begin{equation}
||\mathcal{E}_{\theta}-\mathcal{E}_{\theta}^{M}||_{\diamond}\leq\delta
_{M}:=||\mathcal{I}-\Gamma_{M}||_{\diamond}\leq2\beta M^{-1},
\end{equation}
where we set $\beta:=d(d-1)$. By repeating the steps shown in
Fig.~2 of the main text, we may write the metrological equivalent
of Eq.~(13). In other words, for any input state $\rho_{C}$, we
may write the simulation
\begin{equation}
\mathcal{E}_{\theta}^{M}(\rho_{C})=\mathcal{T}^{M}(\rho_{C}\otimes
\rho_{\mathcal{E}_{\theta}}^{\otimes M}),
\end{equation}
where $\mathcal{T}^{M}$ is a trace-preserving LOCC and $\rho_{\mathcal{E}%
_{\theta}}$ is the Choi matrix of $\mathcal{E}_{\theta}$. Then, we
may also repeat the PBT stretching in Fig.~3 of the main text. In
this way, the $n$-use output state $\rho_{n}=\rho_{n}(\theta)$ of
an adaptive parameter estimation
protocol can be decomposed as in Lemma~2 of the main text, i.e.,%
\begin{equation}
||\rho_{n}(\theta)-\bar{\Lambda}(\rho_{\mathcal{E}_{\theta}}^{\otimes
nM})||\leq n\delta_{M}. \label{lemmaAPP}%
\end{equation}

\subsection{PBT\ implies the Heisenberg scaling}

Using the decomposition in Eq.~(\ref{lemmaAPP}), we may write a
bound for the optimal quantum Fisher information in
Eq.~(\ref{fisheropt}). For large $n$, we obtain the Heisenberg
scaling
\begin{equation}
\overline{\text{\textrm{QFI}}}_{\theta}^{n}\lesssim n^{2}\mathrm{QFI}%
(\rho_{\mathcal{E}_{\theta}}), \label{cube}%
\end{equation}
where
\begin{equation}
\mathrm{QFI}(\rho_{\mathcal{E}_{\theta}})=\frac{4d_{B}^{2}(\rho_{\mathcal{E}%
_{\theta}},\rho_{\mathcal{E}_{\theta+d\theta}})}{d\theta^{2}}.
\end{equation}

In order to show Eq.~(\ref{cube}), consider the function
\begin{equation}
q_{n}(\theta,\delta)=2\frac{d_{B}[\rho_{n}(\theta),\rho_{n}(\theta+\delta
)]}{\delta}.
\end{equation}
We set
$u_{\theta}:=\bar{\Lambda}(\rho_{\mathcal{E}_{\theta}}^{\otimes
nM})$ and apply twice the triangular inequality, so that we may
write
\begin{align}
d_{B}[\rho_{n}(\theta),\rho_{n}(\theta+\delta)]  &  \leq d_{B}[\rho_{n}%
(\theta),u_{\theta}]+\label{cc1}\\
&  d_{B}[u_{\theta},u_{\theta+\delta}]+d_{B}[u_{\theta+\delta},\rho_{n}%
(\theta+\delta)].\nonumber
\end{align}
Bounding the Bures distance with the trace distance, we get
\begin{equation}
d_{B}^{2}[\rho_{n}(\theta),u_{\theta}]\leq\frac{\left\Vert \rho_{n}%
(\theta)-u_{\theta}\right\Vert
}{2}\leq\frac{n\delta_{M}}{2}\leq\frac{\beta
n}{M}. \label{cc2}%
\end{equation}
Using Eqs.~(\ref{cc1}) and~(\ref{cc2}), we may write
\begin{equation}
q_{n}(\theta,\delta)\leq2\frac{d_{B}[u_{\theta},u_{\theta+\delta}]}{\delta
}+\frac{4}{\delta}\sqrt{\frac{\beta n}{M}}. \label{BBC}%
\end{equation}

We may bound $d_{B}$ in Eq.~(\ref{BBC}) as follows%

\begin{align}
&
d_{B}[u_{\theta},u_{\theta+\delta}]\overset{(1)}{\leq}d_{B}[\rho
_{\mathcal{E}_{\theta}}^{\otimes nM},\rho_{\mathcal{E}_{\theta+\delta}%
}^{\otimes nM}]\nonumber\\
&  \overset{(2)}{=}\sqrt{2[1-F(\rho_{\mathcal{E}_{\theta}}^{\otimes nM}%
,\rho_{\mathcal{E}_{\theta+\delta}}^{\otimes nM})]}\nonumber\\
&  \overset{(3)}{=}\sqrt{2(1-F^{nM})}\overset{(4)}{\leq}\sqrt{2nM(1-F)}%
\nonumber\label{eq1}\\
&  \overset{(2)}{=}\sqrt{nM}d_{B}[\rho_{\mathcal{E}_{\theta}},\rho
_{\mathcal{E}_{\theta+\delta}}],
\end{align}
where: (1) we use the monotonicity of the Bures distance under the
CPTP map $\bar{\Lambda}$, (2) we use the standard relation between
Bures distance and
fidelity, (3) we set $F:=F(\rho_{\mathcal{E}_{\theta}},\rho_{\mathcal{E}%
_{\theta+\delta}})$ and exploit the multiplicativity of the
fidelity over tensor products, and (4) we use the inequality
$F^{n}\geq1-n+nF$. Therefore, from Eq.~(\ref{BBC}), we may derive
the inequality
\begin{equation}
q_{n}(\theta,\delta)\leq2\sqrt{nM}\frac{d_{B}[\rho_{\mathcal{E}_{\theta}}%
,\rho_{\mathcal{E}_{\theta+\delta}}]}{\delta}+\frac{4}{\delta}\sqrt
{\frac{\beta n}{M}}.
\end{equation}

Now notice that
\begin{equation}
\lim_{\delta\rightarrow0}2\frac{d_{B}[\rho_{\mathcal{E}_{\theta}}%
,\rho_{\mathcal{E}_{\theta+\delta}}]}{\delta}=\sqrt{\mathrm{QFI}%
(\rho_{\mathcal{E}_{\theta}})}.
\end{equation}
This means that for any $\epsilon>0$, there is
$\delta<\delta_{\epsilon}$ such that
\begin{equation}
q_{n}(\theta,\delta)\leq\sqrt{nM}\left[  \sqrt{\mathrm{QFI}(\rho
_{\mathcal{E}_{\theta}})}+\epsilon\right]
+\frac{4}{\delta}\sqrt{\frac{\beta n}{M}}.
\end{equation}
Setting $M=n^{1+z}$ (for any $z>0$) implies%
\begin{align}
q_{n}(\theta,\delta)  &  \leq\kappa_{n}(\theta,\delta|\epsilon,z)\label{dd1}\\
&  :=\sqrt{n^{2+z}}\left[  \sqrt{\mathrm{QFI}(\rho_{\mathcal{E}_{\theta}}%
)}+\epsilon\right]
+\frac{4}{\delta}\sqrt{\frac{\beta}{n^{z}}}.\nonumber
\end{align}
Note that, by definition, $\mathrm{QFI}_{\theta}(\mathcal{P}_{n}%
):=\lim_{\delta\rightarrow0}q_{n}(\theta,\delta)^{2}$. Then,
assume that the limit
\begin{equation}
\lim_{n\rightarrow\infty}\lim_{\delta\rightarrow0}\frac{q_{n}(\theta
,\delta)^{2}}{n^{2+z}}%
\end{equation}
exists for any $z>0$. Then, using Eq.~(\ref{dd1}), which is valid
for any $n$ and $\delta$, we may write
\begin{align}
\lim_{n\rightarrow\infty}\lim_{\delta\rightarrow0}\frac{q_{n}(\theta,\delta
)}{\sqrt{n^{2+z}}}  &
\leq\underset{n\rightarrow\infty,~\delta\rightarrow
0}{\lim\inf}\frac{\kappa_{n}(\theta,\delta|\epsilon,z)}{\sqrt{n^{2+z}}%
}\nonumber\\
&  \leq\sqrt{\mathrm{QFI}(\rho_{\mathcal{E}_{\theta}})}+\epsilon.
\end{align}
The previous inequality leads to%
\begin{equation}
\lim_{n\rightarrow\infty}\frac{\mathrm{QFI}_{\theta}(\mathcal{P}_{n})}%
{n^{2+z}}\leq\left[  \sqrt{\mathrm{QFI}(\rho_{\mathcal{E}_{\theta}})}%
+\epsilon\right]  ^{2},
\end{equation}
for any $\epsilon,z>0$. Now, sending $\epsilon$ and $z$ to zero
gives the following scaling for large $n$
\begin{equation}
\mathrm{QFI}_{\theta}(\mathcal{P}_{n})\lesssim n^{2}\mathrm{QFI}%
(\rho_{\mathcal{E}_{\theta}})~. \label{cc5}%
\end{equation}
Since this upper bound holds for any protocol $\mathcal{P}_{n}$
(because $\bar{\Lambda}$ disappears), then the asymptotic scaling
in Eq.~(\ref{cc5}) may be extended to
$\overline{\text{\textrm{QFI}}}_{\theta}^{n}$ as in
Eq.~(\ref{cube}). In conclusion we have obtained un upper bound
for the quantum Fisher information corresponding to the Heisenberg
(quadratic) scaling in the number of uses.

\section{Converse bounds for adaptive private communication}

\subsection{Adaptive protocols for quantum/private communication}

Let us assume that the adaptive protocol described in the main
text has the task of secret key generation, i.e., to establish a
secret key between the register $\mathbf{a}$, owned by Alice, and
the register $\mathbf{b}$, owned by Bob. This protocol employs
adaptive LOCCs $\Lambda_{i}$ interleaved with the transmissions
over a $d$-dimensional quantum channel $\mathcal{E}$. (In this
analysis we assume input and output Hilbert spaces with the same
dimension $d$; if the spaces have different dimensions, we may
always pad the one with the lower dimension and formally enlarge
the channel to include the extra dimensions.) After $n$ adaptive
uses of the channel, the output state $\rho_{n}$ of the registers
is epsilon-close to a target private state~\cite{TQCs} $\phi_{n}$
with $nR_{n}^{\epsilon}$ private bits, i.e., $\left\Vert
\rho_{n}-\phi_{n}\right\Vert \leq\epsilon$. By taking the limit
for large $n$, small $\epsilon$\ (weak converse), and optimizing
over all asymptotic key-generation adaptive protocols
$\mathcal{P}$, we define the
secret key capacity of the channel $\mathcal{E}$%
\begin{equation}
K(\mathcal{E}):=\sup_{\mathcal{P}}\lim_{\epsilon,n}R_{n}^{\epsilon}~.
\end{equation}

It is known that this capacity is greater than other two-way
assisted capacities. In fact, we have~\cite{TQCs}
\begin{equation}
Q_{2}(\mathcal{E})=D_{2}(\mathcal{E})\leq P_{2}(\mathcal{E})\leq
K(\mathcal{E}),
\end{equation}
where $Q_{2}$ is the two-way assisted quantum capacity (qubits per
channel use), $D_{2}$ is the two-way assisted entanglement
distribution capacity (ebits per channel use), and $P_{2}$ is the
two-way assisted private capacity (private bits per channel use).
We now investigate upper bounds for $K(\mathcal{E})$ which are
derived by combining PBT stretching with various entanglement
measures, therefore extending one of the main insights of
Ref.~\cite{PLOBs}.

\subsection{PBT stretching of private communication and single-letter upper
bounds}

Consider the $M$-port approximation $\mathcal{E}^{M}$ of
$\mathcal{E}$, as achieved by the PBT simulation with error
$\delta_{M}$. Correspondingly, we have an $M$-port approximate
output state $\rho_{n}^{M}$ such that $\left\Vert
\rho_{n}-\rho_{n}^{M}\right\Vert \leq n\delta_{M}$ as in Eq.~(15)
of the main text. Then, we may stretch an adaptive $n$-use
protocol $\mathcal{P}_{n}$ over
$\mathcal{E}^{M}$ and write $\rho_{n}^{M}=\bar{\Lambda}(\rho_{\mathcal{E}%
}^{\otimes nM})$ for a trace-preserving LOCC $\bar{\Lambda}$.
Using the triangle inequality, we may write
\begin{align}
\left\Vert \rho_{n}^{M}-\phi_{n}\right\Vert  &  \leq\left\Vert \rho_{n}%
^{M}-\rho_{n}\right\Vert +\left\Vert \rho_{n}-\phi_{n}\right\Vert \nonumber\\
&  \leq n\delta_{M}+\epsilon:=\gamma. \label{eqCOMMS}%
\end{align}

Now consider an entanglement measure $E$ with the properties
listed in Ref.~\cite[Sec.~VIII]{TQCs}. For instance, $E$ may be
the relative entropy of entanglement $E_{\text{R}}$
(REE)~\cite{REE1,REE2,REE3} or the squashed entanglement
$E_{\text{SE}}$ (SE). In particular, these measures satisfy a
suitable continuity property. For $d$-dimensional states $\rho$
and $\sigma$ such that $\left\Vert \rho-\sigma\right\Vert
\leq\gamma$, we may write the Fannes-type inequality
\begin{equation}
\left\vert E(\rho)-E(\sigma)\right\vert \leq
g(\gamma)\log_{2}d+h(\gamma),
\label{continuity}%
\end{equation}
where $g$, $h$ are regular functions going to zero in
$\epsilon^{\prime}$. For the REE and the SE, these functions
are~\cite{TQCs}
\begin{align}
\text{REE}  &  \text{:~~}g(\gamma)=4\gamma,~h(\epsilon)=2H_{2}(\gamma),\\
\text{SE}  &  \text{:~~}g(\gamma)=16\sqrt{\gamma},~h(\gamma)=2H_{2}%
(2\sqrt{\gamma}),
\end{align}
where $H_{2}$ is the binary Shannon entropy.

By applying Eq.~(\ref{continuity}) to Eq.~(\ref{eqCOMMS}), we get%
\begin{equation}
\left\vert E(\rho_{n}^{M})-E(\phi_{n})\right\vert \leq
g(\gamma)\log _{2}d+h(\gamma),
\end{equation}
where $E(\phi_{n})\geq nR_{n}^{\epsilon}$ (normalization) and%
\begin{equation}
E(\rho_{n}^{M})=E[\bar{\Lambda}(\rho_{\mathcal{E}}^{\otimes
nM})]\leq nM~E(\rho_{\mathcal{E}}),
\end{equation}
which exploits the monotonicity of $E$ under trace-preserving
LOCCs and the
subadditivity over tensor-product states~\cite{TQCs}. Therefore, we may write%
\begin{equation}
R_{n}^{\epsilon}\leq
M~E(\rho_{\mathcal{E}})+\frac{g(n\delta_{M}+\epsilon
)\log_{2}d+h(n\delta_{M}+\epsilon)}{n}.\label{finakk}%
\end{equation}
Note that for a private state, we may write $\log_{2}d\leq cn$ for
some constant $c$~\cite{TQCs}. Thus, for any adaptive key
generation protocol $\mathcal{P}_{n}$ over a $d$-dimensional
quantum channel $\mathcal{E}$, the maximum $\epsilon$-secure key
rate that can be generated after $n$ uses is bounded as in
Eq.~(\ref{finakk}) where $E$ is an entanglement measure (as the
REE or the SE), $M$ is the number of ports, and $\delta_{M}$ is
the error of the $M$-port PBT defined in Lemma~1 of the main text.

We can find alternate bound by extending the definition of
channel's REE~\cite{PLOBs} to a tripartite version. Consider three
finite-dimensional systems $a^{\prime}$, $a$ and $b^{\prime}$, and
a quantum channel $\mathcal{E}=\mathcal{E}_{a\rightarrow b}$ from
$a$ to the output system $b$. Consider a generic input state
$\rho_{a^{\prime}ab^{\prime}}$ transformed into an output state
$\omega_{_{a^{\prime}bb^{\prime}}}:=\mathcal{E}_{a\rightarrow
b}(\rho_{a^{\prime}ab^{\prime}})$ by the action of this channel.
Then, one can
define a tripartite version of channel's REE as%
\begin{equation}
\tilde{E}_{\text{R}}(\mathcal{E}):=\sup_{\rho_{a^{\prime}ab^{\prime}}%
}E_{\text{R}}(a^{\prime}|bb^{\prime})_{\omega}-E_{\text{R}}(a^{\prime
}a|b)_{\rho},
\end{equation}
which satisfies
$K(\mathcal{E})\leq\tilde{E}_{\text{R}}(\mathcal{E})$~\cite{Kaur}.
Moreover, if two channels are close in diamond norm $\Vert\mathcal{E}%
-\mathcal{E}^{\prime}\Vert_{\diamond}\leq2\epsilon$, then one may
also write the continuity property~\cite{Kaur}
\begin{gather}
|\tilde{E}_{\text{R}}(\mathcal{E})-\tilde{E}_{\text{R}}(\mathcal{E}^{\prime
})|\leq2\epsilon\log_{2}{d}+f(\epsilon),\label{gg1}\\
f(\epsilon):=(1+\epsilon)\log_{2}{(1+\epsilon)}-\epsilon\log_{2}{\epsilon,}%
\end{gather}
where $d$ is the dimension of the Hilbert space. Finally, as a
straightforward application of one of the tools established in
Ref.~\cite{PLOBs}, i.e., the LOCC simulation of a quantum channel
$\mathcal{E}$ via a resource state $\sigma$~\cite{TQCs}, one may
write the data-processing upper bound $\tilde
{E}_{\text{R}}(\mathcal{E})\leq E_{\text{R}}(\sigma)$.

In our channel simulation via PBT, we have a multi-copy resource
state $\sigma=\rho_{\mathcal{E}}^{\otimes M}$ for the $M$-port
approximation $\mathcal{E}^{M}$ of the $d$-dimensional channel
$\mathcal{E}$. This means that we may write
\begin{equation}
\tilde{E}_{\text{R}}(\mathcal{E}^{M})\leq E_{\text{R}}(\rho_{\mathcal{E}%
}^{\otimes M})\leq ME_{\text{R}}(\rho_{\mathcal{E}}).
\end{equation}
Then, because we have%
\begin{equation}
||\mathcal{E}-\mathcal{E}^{M}||_{\diamond}\leq||\mathcal{I}-\Gamma
_{M}||_{\diamond}:=\delta_{M}\leq2d(d-1)M^{-1},
\end{equation}
from Eq.~(\ref{gg1}) we may derive%
\begin{equation}
\tilde{E}_{\text{R}}(\mathcal{E})\leq
E_{\text{R}}(\rho_{\mathcal{E}}^{\otimes
M})+\delta_{M}\log_{2}{d}+f(\delta_{M}/2).
\end{equation}
As a result, we may write the upper bound%
\begin{align}
K(\mathcal{E})  &  \leq E_{\text{R}}(\rho_{\mathcal{E}}^{\otimes M}%
)+\delta_{M}\log_{2}{d}+f(\delta_{M}/2)\nonumber\\
&  \leq ME_{\text{R}}(\rho_{\mathcal{E}})+\frac{2d(d-1)}{M}\log_{2}%
{d}+f\left[  \frac{d(d-1)}{M}\right] \nonumber\\
&  :=K_{\text{UB}}^{M}(\mathcal{E}).
\end{align}
The tightest upper bound is obtained by minimizing $K_{\text{UB}}%
^{M}(\mathcal{E})$ over $M$, which is typically a finite value.

Let us apply the bound to channels that are nearly
entanglement-breaking, so that
$E_{\text{R}}(\rho_{\mathcal{E}})\ll1$. In this case, we expect
that the optimal value of $M$ is large. It is easy to see that a
sub-optimal choice for
$M$ is given by%
\begin{equation}
\tilde{M}=\sqrt{\frac{2d(d-1)\log_{2}{d}}{E_{\text{R}}(\rho_{\mathcal{E}})}%
}\,,
\end{equation}
which provides the upper bound
\begin{align}
K(\mathcal{E})  &  \leq2\sqrt{2d(d-1)\log_{2}{d}}\sqrt{E_{\text{R}}%
(\rho_{\mathcal{E}})}\nonumber\\
&  +f\left[  \sqrt{\frac{d(d-1)E_{\text{R}}(\rho_{\mathcal{E}})}{2\log_{2}{d}%
}}\right]  \,. \label{boundDD}%
\end{align}
The bound in Eq.~(\ref{boundDD}) is particularly interesting for
almost
entanglement-breaking channels, such that $E_{\text{R}}(\rho_{\mathcal{E}%
})\lesssim(\log_{2}d)/[8d(d-1)]$.


\begin{thebibliography}{99}                                                                                               %


\bibitem {QHT1}Helstrom, C. W. \textit{Quantum Detection and Estimation
Theory} (New York: Academic, 1976).


\bibitem {Watrous}Watrous, J. \textit{The theory of quantum information}
(Cambridge Univ. Press, Cambridge, 2018).

\bibitem {HolevoBOOK}Holevo, A.\textit{ Quantum Systems, Channels,
Information: A Mathematical Introduction} (De Gruyter, Berlin,
2012).

\bibitem {NiCh}Nielsen, M. A. \& Chuang, I. L. \textit{Quantum computation and
quantum information} (Cambridge Univ. Press, Cambridge, 2000).

\bibitem {RMP}Weedbrook, C. \textit{et al.} Gaussian quantum information.
\textit{Rev. Mod. Phys.} \textbf{84}, 621 (2012).

\bibitem {QCB1}Audenaert, K. M. R. \textit{et al.} Discriminating States: The
Quantum Chernoff Bound. \textit{Phys. Rev. Lett.} \textbf{98}, 160501 (2007).

\bibitem {QCB2}Calsamiglia, J., Munoz-Tapia, R., Masanes, L., Acin, A. \&
Bagan, E. The quantum Chernoff bound as a measure of distinguishability
between density matrices: application to qubit and Gaussian states.
\textit{Phys. Rev. A} \textbf{77}, 032311 (2008).

\bibitem {QCB3}Pirandola, S. \& Lloyd, S. Computable bounds for the
discrimination of Gaussian states. \textit{Phys. Rev. A} \textbf{78}, 012331 (2008).

\bibitem {QHB1}Audenaert, K. M. R., Nussbaum, M., Szkola, A. \& Verstraete, F.
Asymptotic Error Rates in Quantum Hypothesis Testing. \textit{Commun. Math.
Phys.} \textbf{279}, 251 (2008).

\bibitem {QHB2}Spedalieri, G. \& Braunstein, S. L. Asymmetric quantum
hypothesis testing with Gaussian states. \textit{Phys. Rev. A} \textbf{90},
052307 (2014).

\bibitem {QCD1}Acin, A. Statistical distinguishability between unitary
operations. \textit{Phys. Rev. Lett.} \textbf{87}, 177901 (2001).

\bibitem {QCD2}Sacchi, M. Entanglement can enhance the distinguishability of
entanglement-breaking channels. \textit{Phys. Rev. A} \textbf{72}, 014305 (2005).

\bibitem {QCD3}Wang, G. \& Ying, M. Unambiguous discrimination among quantum
operations. \textit{Phys. Rev. A} \textbf{73}, 042301 (2006).

\bibitem {QCD4}Childs, A., Preskill, J. \& Renes, J. Quantum information and
precision measurement. \textit{J. Mod. Opt.} \textbf{47}, 155 (2000).

\bibitem {QCD6}Invernizzi, C., Paris, M. G. A. \& Pirandola, S. Optimal
detection of losses by thermal probes. \textit{Phys. Rev. A} \textbf{84},
022334 (2011).

\bibitem {HayaCLASS}Hayashi, M. Discrimination of two channels by
adaptive methods and its application to quantum system.
\textit{IEEE Trans. Inf. Theory} \textbf{55}, 3807 (2009).

\bibitem {PirCo}Pirandola, S. \& Lupo, C. Ultimate precision of adaptive noise
estimation. \textit{Phys. Rev. Lett.} \textbf{118}, 100502 (2017).



\bibitem {ReviewMETRO}Pirandola, S., Bardhan, B. R., Gehring, T.,
Weedbrook, C. \& Lloyd, S. Advances in Photonic Quantum Sensing.
\textit{Nat. Photon.} \textbf{12}, 724-733 (2018).

\bibitem {Harrow}Harrow, A. W., Hassidim, A., Leung, D. W. \& Watrous, J.
Adaptive versus non-adaptive strategies for quantum channel discrimination.
\textit{Phys. Rev. A} \textbf{81}, 032339 (2010).

\bibitem {Diamond}Paulsen, V. I. \textit{Completely Bounded Maps and Operator
Algebras} (Cambridge Univ. Press, Cambridge, 2002).

\bibitem {PBT}Ishizaka, S. \& Hiroshima, T. Asymptotic teleportation scheme as
a universal programmable quantum processor. \textit{Phys. Rev. Lett.}
\textbf{101}, 240501 (2008).

\bibitem {PBT1}Ishizaka, S. \& Hiroshima, T. Quantum teleportation scheme by
selecting one of multiple output ports. \textit{Phys. Rev. A} \textbf{79},
042306 (2009).

\bibitem {PBT2}Ishizaka, S. Some remarks on port-based teleportation. Preprint
at https://arxiv.org/abs/1506.01555 (2015).

\bibitem {Brau}Wang, Z.-W. \& Braunstein, S. L. Higher-dimensional performance
of port-based teleportation. \textit{Sci. Rep.} \textbf{6}, 33004 (2016).

\bibitem {PLOB}Pirandola, S., Laurenza, R., Ottaviani, C. \& Banchi, L.
Fundamental limits of repeaterless quantum communications. \textit{Nat.
Commun.} \textbf{8}, 15043 (2017). See also preprint at
https://arxiv.org/abs/1510.08863 (2015).

\bibitem {TQC}Pirandola, S., Braunstein, S. L., Laurenza, R., Ottaviani, C.,
Cope, T. P. W., Spedalieri, G. \& Banchi, L. Theory of channel simulation and
bounds for private communication \textit{Quantum Sci. Technol.} \textbf{3},
035009 (2018).

\bibitem {Uniform}Pirandola, S., Laurenza, R. \& Braunstein, S. L.
Teleportation simulation of bosonic Gaussian channels: Strong and uniform
convergence. Eur. Phys. J. D \textbf{72}, 162 (2018).

\bibitem {Qill0}Lloyd, S. Enhanced sensitivity of photodetection via quantum
illumination. \textit{Science} \textbf{321}, 1463 (2008).

\bibitem {Qill1}Tan, S.-H. \textit{et al.} Quantum illumination with Gaussian
states. \textit{Phys. Rev. Lett.} \textbf{101}, 253601 (2008).

\bibitem {Qill4}Shapiro, J. H. \& Lloyd, S. Quantum illumination versus
coherent-state target detection. \textit{New J. Phys.} \textbf{11}, 063045 (2009).

\bibitem {Qill6}Zhang, Z., Tengner, M., Zhong, T., Wong, F. N. C. \& Shapiro,
J. H. Entanglement's benefit survives an entanglement-breaking channel.
\textit{Phys. Rev. Lett.} \textbf{111}, 010501 (2013).

\bibitem {Qill7}Lopaeva, E. D., Ruo Berchera, I., Degiovanni, I. P., Olivares,
S., Brida, G. \& Genovese, M. Experimental realization of quantum
illumination. \textit{Phys. Rev. Lett.} \textbf{110}, 153603 (2013).

\bibitem {Qill8}Zhang, Z., Mouradian, S., Wong, F. N. C. \& Shapiro, J. H.
Entanglement-enhanced sensing in a lossy and noisy environment. \textit{Phys.
Rev. Lett.} \textbf{114}, 110506 (2015).

\bibitem {Qill2}Barzanjeh, S. \textit{et al. }Microwave quantum illumination.
\textit{Phys. Rev. Lett.} \textbf{114}, 080503 (2015).

\bibitem {Qill3}Weedbrook, C., Pirandola, S., Thompson, J., Vedral, V. \& Gu,
M. How discord underlies the noise resilience of quantum illumination.
\textit{New J. Phys.} \textbf{18}, 043027 (2016).


\bibitem {Array}Nielsen, M. A. \& Chuang, I. L. Programmable quantum gate
arrays. \textit{Phys. Rev. Lett.} \textbf{79}, 321 (1997).

\bibitem {networkPIRS}Pirandola, S. End-to-end capacities of a quantum communication network.
\textit{Commun. Phys.} \textbf{2}, 51 (2019). See also preprint at
https://arxiv.org/abs/1601.00966 (2016).

\bibitem {Multipoint}Laurenza, R. \& Pirandola, S. General bounds for
sender-receiver capacities in multipoint quantum communications. \textit{Phys.
Rev. A} \textbf{96}, 032318 (2017).

\bibitem {finiteStretching}Laurenza, R., Braunstein, S. L. \& Pirandola, S.
Finite-resource teleportation stretching for continuous-variable
systems. \textit{Sci. Rep.} \textbf{8}, 15267 (2018). See also
preprint at https://arxiv.org/abs/1706.06065 (2017).

\bibitem {nonPauli}Cope, T. P. W., Hetzel, L., Banchi, L. \& Pirandola, S.
Simulation of non-Pauli channels. \textit{Phys. Rev. A} \textbf{96}, 022323 (2017).

\bibitem {HWchannels}Cope, T. P. W. \& Pirandola, S. Adaptive estimation and
discrimination of Holevo-Werner channels. \textit{Quantum Meas. Quantum
Metrol.} \textbf{4}, 44-52 (2017).

\bibitem {Fuchs}Fuchs, C. A. \& van de Graaf, J. Cryptographic
distinguishability measures for quantum mechanical states. \textit{IEEE Trans.
Inf. Theory} \textbf{45,} 1216 (1999).

\bibitem {Pin1}Pinsker, M. S. Information and Information Stability of
Random Variables and Processes (San Francisco, Holden Day, 1964).

\bibitem {Pin2}Carlen, E. A. \& Lieb, E. H. Bounds for entanglement
via an extension of strong subadditivity of entropy. \textit{Lett.
Math. Phys.} \textbf{101}, 1-11 (2012).

\bibitem {Tsang15}Tsang, M., Nair, R. \& Lu, X.-M. Quantum theory of
superresolution for two incoherent optical point sources. \textit{Phys. Rev.
X} \textbf{6}, 031033 (2016).

\bibitem {Lupo16}Lupo, C. \& Pirandola, S. Ultimate precision bound of quantum
and subwavelength imaging. \textit{Phys. Rev. Lett.} \textbf{117}, 190802 (2016).

\bibitem {Tsang2}Nair, R. \& Tsang, M. Far-Field Superresolution of thermal
electromagnetic sources at the quantum limit. \textit{Phys. Rev. Lett.}
\textbf{117}, 190801 (2016).


\bibitem {dd3}Cooney, T., Mosonyi, M. \& Wilde, M. M. Strong converse
exponents for a quantum channel discrimination problem and
quantum-feedback-assisted communication. \textit{Comm. Math.
Phys.} \textbf{344}, 797-829 (2016).

\bibitem {Qill5}De Palma, G. \& Borregaard, J. The minimum error probability
of quantum illumination. \textit{Phys. Rev. A} \textbf{98}, 012101
(2018).




\bibitem {Qread}Pirandola, S. Quantum reading of a classical digital memory.
\textit{Phys. Rev. Lett.} \textbf{106}, 090504 (2011).

\bibitem {QreadCAP}Pirandola, S., Lupo, C., Giovannetti, V., Mancini, S. \&
Braunstein, S. L. Quantum reading capacity. \textit{New J. Phys.} \textbf{13},
113012 (2011).

\bibitem {Arno12}Dall'Arno, M., Bisio, A., D'Ariano, G. M., Mikov\'{a}, M.,
Je\v{z}ek, M. \& Du\v{s}ek, M. Experimental implementation of unambiguous
quantum reading. \textit{Phys. Rev. A} \textbf{85}, 012308 (2012).

\bibitem {ArnoIJQI}Dall'Arno, M., Bisio, A. \& D'Ariano, G. M. Ideal quantum
reading of optical memories. \textit{Int. J. Quant. Inf.} \textbf{10}, 1241010 (2012).

\bibitem {GaeENTROPY}Spedalieri, G. Cryptographic aspects of quantum reading.
\textit{Entropy} \textbf{17}, 2218-2227 (2015).

\bibitem {Sam1}Braunstein, S. L. \& Caves, C. M. Statistical distance and the
geometry of quantum states. \textit{Phys. Rev. Lett.} \textbf{72}, 3439 (1994).

\bibitem {Sam2}Braunstein, S. L., Caves, C. M. \& Milburn, G. J. Generalized
uncertainty relations: theory, examples, and Lorentz invariance. \textit{Ann.
Phys.} \textbf{247}, 135-173 (1996).

\bibitem {Paris}Paris, M. G. A. Quantum estimation for quantum technology.
\textit{Int. J. Quant. Inf.} \textbf{7}, 125-137 (2009).

\bibitem {Giova}Giovannetti, V., Lloyd, S. \& Maccone, L. Advances in quantum
metrology. \textit{Nature Photon.} \textbf{5}, 222 (2011).

\bibitem {ReviewNEW}Braun, D. \textit{et al.} Quantum enhanced measurements
without entanglement. \textit{Rev. Mod. Phys.} \textbf{90}, 035006 (2018).

\bibitem {Doukas}Doukas, J., Adesso, G., Pirandola, S. \& Dragan, A.
Discriminating quantum field theories in non-inertial frames. \textit{Class.
Quantum Grav.} \textbf{32}, 035013 (2015).

\bibitem {Majenz}Majenz, C. Entropy in Quantum Information Theory,
Communication and Cryptography (PhD thesis, University of
Copenhagen, 2017).

\bibitem {nechita}Nechita, I. \textit{et al.} Almost all quantum channels are
equidistant. \textit{J. of Math. Phys.} \textbf{59}, 052201
(2018).
\end{thebibliography}

\begin{thebibliography}{99}                                                                                               %


\bibitem {PBT1s}Ishizaka, S. \& Hiroshima, T. Quantum teleportation scheme by
selecting one of multiple output ports. \textit{Phys. Rev. A}
\textbf{79}, 042306 (2009).

\bibitem {Shap}Shapiro, J. H. The quantum theory of optical communications.
\textit{IEEE J. Sel. Top. Quantum Electron.} \textbf{15},
1547-1569 (2009).

\bibitem {m1}Lupo, C., Giovannetti, V., Pirandola, S., Mancini, S. \& Lloyd,
S. Enhanced quantum communication via optical refocusing.
\textit{Phys. Rev. A }\textbf{84}, 010303(R) (2011).

\bibitem {m2}Lupo, C., Giovannetti, V., Pirandola, S., Mancini, S. \& Lloyd,
S. Capacities of linear quantum optical systems. \textit{Phys.
Rev. A} \textbf{85}, 062314 (2012).

\bibitem {NiChs}Nielsen, M. A. \& Chuang I.\ L. \textit{Quantum computation and
quantum information} (Cambridge University Press, Cambridge,
2000).

\bibitem {Tsang15s}Tsang, M., Nair, R \& Lu, X.-M. Quantum Theory of
Superresolution for Two Incoherent Optical Point Sources.\
\textit{Phys. Rev. X} \textbf{6}, 031033 (2016).

\bibitem {Lupo16s}Lupo, C. \& Pirandola, S. Ultimate Precision Bound of Quantum
and Subwavelength Imaging. \textit{Phys. Rev. Lett.} \textbf{117},
190802 (2016).

\bibitem {Qill0s}Lloyd, S. Enhanced sensitivity of photodetection via quantum
illumination. \textit{Science} \textbf{321}, 1463 (2008).

\bibitem {Qill1s}Tan, S.-H. \textit{et al.} Quantum illumination with Gaussian
states. \textit{Phys. Rev. Lett.} \textbf{101}, 253601 (2008).

\bibitem {Qill2s}Barzanjeh, S \textit{et al.} Microwave quantum illumination.
\textit{Phys. Rev. Lett.} \textbf{114}, 080503 (2015).

\bibitem {Qill3s}Weedbrook, C., Pirandola, S., Thompson, J., Vedral, V. \& Gu,
M. How discord underlies the noise resilience of quantum
illumination. \textit{New J. Phys. }\textbf{18}, 043027 (2016).

\bibitem {Sam1s}Braunstein, S. L. \& Caves, C. M. Statistical distance and the
geometry of quantum states. \textit{Phys. Rev. Lett.} \textbf{72},
3439 (1994).

\bibitem {PirCos}Pirandola, S. \& Lupo, C. Ultimate precision of adaptive noise
estimation. \textit{Phys. Rev. Lett.} \textbf{118}, 100502 (2017).

\bibitem {ReviewMETROs}Laurenza, R., Lupo, C., Spedalieri, G., Braunstein, S.
L. \& Pirandola, S. Channel simulation in quantum metrology.
\textit{Quantum Meas. Quantum Metrol.} \textbf{5}, 1-12 (2018).

\bibitem {TQCs}Pirandola, S., Braunstein, S. L., Laurenza, R., Ottaviani, C.,
Cope, T. P. W., Spedalieri, G. \& Banchi, L. Theory of channel
simulation and bounds for private communication \textit{Quantum
Sci. Technol.} \textbf{3}, 035009 (2018).

\bibitem {PLOBs}Pirandola, S., Laurenza, R., Ottaviani, C. \& Banchi, L.
Fundamental limits of repeaterless quantum communications.
\textit{Nat. Commun.} \textbf{8}, 15043 (2017). See also preprint
at https://arxiv.org/abs/1510.08863 (2015).

\bibitem {REE1}Vedral, V. The role of relative entropy in quantum information
theory. \textit{Rev. Mod. Phys. }\textbf{74}, 197 (2002).

\bibitem {REE2}Vedral, V., Plenio, M. B., Rippin, M. A. \& Knight, P. L.
Quantifying entanglement. \textit{Phys. Rev. Lett. }\textbf{78},
2275-2279 (1997).

\bibitem {REE3}Vedral, V. \& Plenio, M. B. Entanglement measures and
purification procedures. \textit{Phys. Rev. A} \textbf{57}, 1619
(1998).

\bibitem {Kaur}Kaur, E. \& Wilde, M. M. Amortized entanglement of a quantum channel and approximately
teleportation-simulable channels. \textit{J. of Phys. A}
\textbf{51}, 035303 (2018).




\end{thebibliography}
\end{document}